\documentclass{amsart}
\usepackage{cite}

\usepackage{amsfonts} %para el \mathbb
\usepackage{algorithmic} %para los algoritmos
\usepackage{multirow}
\usepackage{multicol}
\usepackage{array}
\usepackage{cprotect}%para usar \cprotect en los captions y poder usar \verb en ellos
\usepackage{color}
\usepackage{tikz}

\usepackage{array}

\usepackage[hidelinks]{hyperref}
\usepackage{url}
\usepackage{mathtools,amsmath,amssymb,amsthm} % For including math equations, theorems, symbols, etc
\usepackage{amsfonts} %para el mathbb
\usepackage{color}
\usepackage{soul}%para el tachado
\usepackage{mathtools,amsmath,amssymb,amsthm} % For including math equations, theorems, symbols, etc
\usepackage[overload]{empheq}

\newtheorem{definition}{Definition}[section]
\theoremstyle{remark}
\newtheorem*{remark}{Remark}

\begin{document}

\newcommand{\bluesolidline}{\raisebox{2pt}{\tikz{\draw[-,blue,solid,line width = 0.7pt](0,0) -- (5mm,0);}}}
\newcommand{\bluedashedline}{\raisebox{2pt}{\tikz{\draw[-,blue,dashed,line width = 0.7pt](0,0) -- (5mm,0);}}}
\newcommand{\bluedasheddottedline}{\raisebox{2pt}{\tikz{\draw[-,blue,dash dot,line width = 0.7pt](0,0) -- (5mm,0);}}}

\newcommand{\magentasolidline}{\raisebox{2pt}{\tikz{\draw[-,magenta,solid,line width = 0.7pt](0,0) -- (5mm,0);}}}
\newcommand{\magentadashedline}{\raisebox{2pt}{\tikz{\draw[-,magenta,dashed,line width = 0.7pt](0,0) -- (5mm,0);}}}
\newcommand{\magentadasheddottedline}{\raisebox{2pt}{\tikz{\draw[-,magenta,dash dot,line width = 0.7pt](0,0) -- (5mm,0);}}}

\newcommand{\redsolidline}{\raisebox{2pt}{\tikz{\draw[-,red,solid,line width = 0.7pt](0,0) -- (5mm,0);}}}
\newcommand{\blacksolidline}{\raisebox{2pt}{\tikz{\draw[-,black,solid,line width = 0.7pt](0,0) -- (5mm,0);}}}
\newcommand{\cyansolidline}{\raisebox{2pt}{\tikz{\draw[-,cyan,solid,line width = 0.7pt](0,0) -- (5mm,0);}}}

\title[Decomposing complicated signals]{Decomposing non-stationary signals with time-varying wave-shape functions}


\author{Marcelo A. Colominas and Hau-Tieng Wu}
\thanks{M. A. Colominas is with the Institute for Research and Development in Bioengineering and Bioinformatics (IBB), CONICET, Ruta Prov. 11 Km. 10, Oro Verde, Entre R\'ios, Argentina (e-mail: macolominas@conicet.gov.ar).}% <-this % stops a space
\thanks{H.-T. Wu is  with the Department of Mathematics and Department of Statistical Science, Duke University, Durham, NC, 27708 USA (email: hauwu@math.duke.edu).}% <-this % stops a space
\begin{abstract}
Modern time series are usually composed of multiple oscillatory components, with time-varying frequency and amplitude contaminated by noise. The signal processing mission is further challenged if each component has an oscillatory pattern, or the wave-shape function, far from a sinusoidal function, and the oscillatory pattern is even changing from time to time. In practice, if multiple components exist, it is desirable to robustly decompose the signal into each component for various purposes, and extract desired dynamics information. Such challenges have raised a significant amount of interest in the past decade, but a satisfactory solution is still lacking. We propose a novel {\em nonlinear regression scheme} to robustly decompose a signal into its constituting multiple oscillatory components with time-varying frequency, amplitude and wave-shape function. We coined the algorithm {\em shape-adaptive mode decomposition (SAMD)}. In addition to simulated signals, we apply SAMD to two physiological signals, impedance pneumography and electroencephalography. Comparison with existing solutions, including linear regression, recursive diffeomorphism-based regression and multiresolution mode decomposition, shows that our proposal can provide an accurate and meaningful decomposition with computational efficiency.
\\
\\
\noindent \textbf{Keywords:} wave-shape functions, signal modeling, instantaneous frequency, biomedical signals
\end{abstract}

\maketitle

% \fi
%\IEEEpeerreviewmaketitle

\section{Introduction}

Modeling real-world oscillatory signals in a compact and physically meaningful manner and developing a suitable algorithm to analyze such signals remain a challenging topic in signal processing. Among various challenges, one shared by many real-world signals is that the amplitude and frequency of each oscillatory component is time-varying, and the oscillatory pattern is usually not sinusoidal. With the advance of sensor technology, examples can be found in various areas, such as biomedicine \cite{akay1998time,Colominas2014,wu2020current}, physics \cite{chassande2016ondelettes,Pham2017high}, to name but a few.

One model that has been widely considered in the scientific community in the last decades is the superimposition of amplitude- and frequency-modulated (AM-FM) components.
We call a signal satisfying the following condition an {\em intrinsic mode type (IMT)} function:
\begin{align}\label{Introduction:ANHM0}
A(t)\cos(2\pi\phi(t))\,,
\end{align}
where $t\in \mathbb{R}$, $A\in C^1(\mathbb{R})$ is a positive smooth function indicating the time-varying amplitude called the ``amplitude modulation'' (AM),
$\phi\in C^2(\mathbb{R})$ is a smooth and monotonically increasing function quantifying how fast the signal oscillates called the {\em phase function}, %
and $\phi'$ is a positive function called the {\em instantaneous frequency (IF)}.
If a function can be represented as a superimposition of multiple IMT functions with some growth conditions and possibly noises, we say that it satisfies the adaptive harmonic model (AHM) \cite{Daubechies2011synchro,Chen_Cheng_Wu:2014}. Other models are also possible, for example, the famous analytic signal model \cite{Gabor:1946}, the time-varying autoregressive model \cite{abramovich2007time}, the TBATS (Trigonometric seasonality, Box-Cox transformation, ARIMA errors, Trend and Seasonal components) model \cite{de2011forecasting}, or the wave-shape oscillatory model \cite{lin2019wave}. We mention that the analytic model is recently extensively studied via the complex analysis perspective \cite{Nahon:2000Thesis}.

While the AHM and other models have been widely applied, they usually ignore a critical aspect -- in many real signals, the waveform morphology can be more complicated than a sinusoidal oscillation. We could model the oscillatory pattern by the wave-shape function (WSF) $s(t)$ \cite{HTWu2013}, which is a $1$-periodic function, i.e. $\min_{\tau>0}\{s(t) = s(t + \tau)|\,t\in\mathbb{R}\}=1$. For example, the function $\cos(2\pi\cdot)$ in \eqref{Introduction:ANHM0} is the typical WSF that is sinusoidal. However, in general the oscillatory pattern is far from being sinusoidal. See Fig. \ref{fig:01} for two example signals, including one electrocardiogram (ECG) and one respiratory signal.
Such complicated non-sinusoidal oscillation usually encodes significant information for practitioners. For example, physicians identify myocardial infarction by reading if the ST segment is elevated from the electrocardiograph signal. See Fig. \ref{fig:01} (a). From this example, we also see that the WSF is not fixed but changes with time \cite{CYLin2018}. Another example is that the inspiration expiration ratio (IER) \cite{strauss2000relative} changes from cycle to cycle, which reflects and also impacts the physiological systems. See Fig. \ref{fig:01} (b) for an illustration of IER. If a signal is the superimposition of multiple AM-FM components with non-sinusoidal oscillation, that is, the $\cos(2\pi\cdot)$ in an IMT function is replaced by a non-sinusoidal function, we say that the signal satisfies the {\em adaptive non-harmonic model} (ANHM). Mathematical details will be provided below. See Fig. \ref{fig:01} (b) for an example that is composed of two non-sinusoidal oscillatory components; one is the respiratory component, and one is the cardiac component usually called the cardiogenic artifact. In the past few years, how to model and analyze this kind of signal has attracted a lot of attention \cite{tavallali2014extraction,Pahlevan_Tavallali_Rinderknecht_Petrasek_Matthews_Hou_Gharib:2014,hou2016extracting,Yang2019multiresolution,Xu2018recursive}, from both theoretical and algorithmic perspectives.

\begin{figure}[t]
\begin{center}
(a) \includegraphics[width=0.95\textwidth]{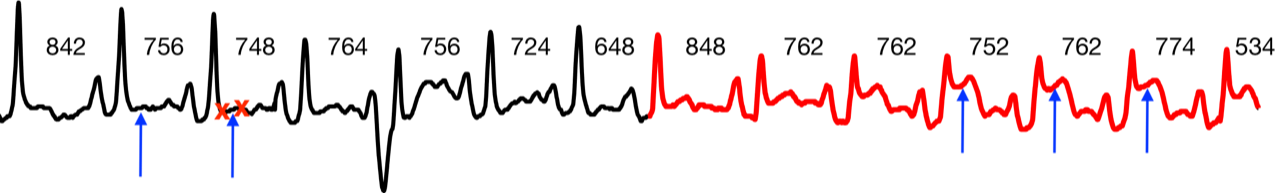}\\
(b) \includegraphics[width=0.95\textwidth]{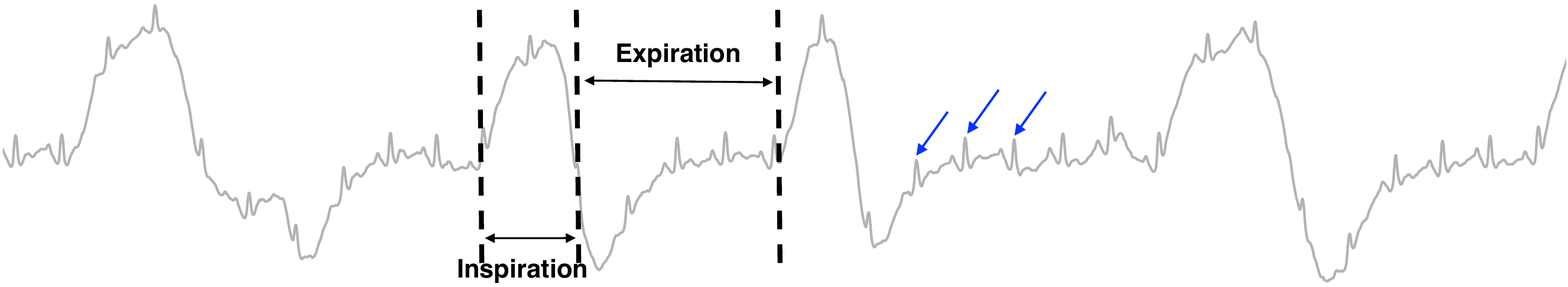}
\end{center}
\caption{An illustration of two complicated non-sinusoidal oscillatory signals, an electrocardiogram (ECG) (a) and a respiratory signal (b). In (a), the ECG signal suggests a migration into myocardial infarction (colored by red). Two red crosses on the left hand side indicates a ST segment. Clearly, the ST segments indicated by blue arrows elevate in the second half. The numbers are the length of R peak to R peak intervals, with the unit ms. In (b), the inspiration and expiration periods are marked by dashed vertical lines. The three blue arrows indicate the second oscillatory component, the cardiogenic artifact.}
\label{fig:01}
\end{figure}

With the ANHM, there are several common signal processing missions, including but not exclusively, how to estimate the IF, AM and the phase, how to extract the WSF of each oscillatory component, how to decompose the signal into each oscillatory component, how to achieve the above missions robustly, efficiently, or even in real time, etc.
There have been several solutions toward these missions. For example, de-shape algorithm was proposed to estimate IF \cite{CYLin2018}; the synchrosqueezing transform (SST) was applied to extract phase \cite{huang2020scientific}; the manifold learning idea was applied to estimate the WSF \cite{lin2019wave};
the linear regression \cite{Wu2016modeling} and optimization approaches \cite{hou2016extracting,Yang2019multiresolution,Xu2018recursive} were proposed to decompose signals. Among these missions, in general the signal decomposition mission so far remains a challenging one, particularly when the WSF is complicated.

In this paper, we focus on the signal decomposition mission. We propose a new approach to robustly decompose signals composed of time-varying WSF into its components. It is motivated by noticing the success of the linear regression approach \cite{Wu2016modeling}, and its limitation coming from the fixed WSF assumption. In our approach, which we coined {\em Shape-adaptive mode decomposition} (SAMD), we generalize the linear regression approach to a nonlinear regression problem to accommodate time-varying WSFs. The design of the algorithm balances between the decomposition accuracy and computational complexity.

The rest of the paper is organized as follows. In Sec. \ref{sec:Models}, we recall the basic concepts of WSF giving its definition, along with a model for time-varying WSFs, and we clearly state the problem we are tackling. In Sec. \ref{sec:Algorithms} we review existing algorithms and introduce our nonlinear regression approach. Numerical results are presented in Sec. \ref{sec:Results}, where we analyze simulated signals and offer two real-world examples by processing electroencephalography and impedance pneumography signals. Sec. \ref{sec:Conclusions} concludes the paper. More numerical results, including a three-component signal decomposition, can be found in the Supplemental Material.

\section{Models}\label{sec:Models}
Before introducing our proposed algorithm, we summarize the ANHM with time-varying WSF.

\subsection{Wave-shape Functions}
We start with the traditional model \cite{HTWu2013}. It was proposed to model (trend- and noise-free) oscillatory signals as a pair amplitude-oscillation of the form,
\begin{equation}\label{eq:WSF}
f_{\texttt{fix}}(t) = A(t) s(\phi(t))\,,
\end{equation}
where $A(t)$ and $\phi(t)$ are the same as those shown in \eqref{Introduction:ANHM0}, and $s(t)$ is a $1$-periodic WSF. We need some conditions for \eqref{eq:WSF}. First, the model in \eqref{eq:WSF} must fulfill the following \emph{slowly varying} conditions \cite{HTWu2013}:
\begin{itemize}
\item[C1.] For $\epsilon >0$, we have $|A'(t)| \leq \epsilon \phi'(t)$ and $|\phi''(t)|\leq \epsilon \phi'(t)$ for all $t \in \mathbb{R}$.
\item[C2.] $\| \phi'' \|_{\infty} \leq M$, with $M \geq 0$.
\end{itemize}
To put some conditions on $s(t)$, note that its Fourier series satisfies $s(t) = \sum_{\ell \in \mathbb{Z}} \hat{s}(\ell) e^{i\ell2\pi t}$, where $\hat{s}(\ell)$ are the Fourier coefficients. In \cite{HTWu2013}, a class of functions was considered:

\begin{definition}\label{def:WSF}
\textbf{Analytic shape function class $\mathcal{S}^{\delta,D,\theta}$. \cite{HTWu2013}} Given $\delta \geq 0$, $D \in \mathbb{N}$, and $\theta \geq 0$, the class $\mathcal{S}^{\delta,D,\theta}$ is defined as the $1$-periodic functions of zero-mean ($\hat{s}(0) = 0$), with unit $L^2$-norm, and satisfying:
\begin{enumerate}
\item[S1.] for all $ \ell \in \mathbb{Z}$, with $|\ell| \neq 1$, $|\hat{s}(\ell)| \leq \delta|\hat{s}(1)|$
\item[S2.] $\sum_{|\ell|>D} |\ell\hat{s}(\ell)|\leq \theta$
\end{enumerate}
\end{definition}
The parameters $\delta$, $D$ and $\theta$ characterize the ``shape'' of the function. The condition S1 says that the strength of the fundamental component  cannot be zero. The condition S2 says that the shape does not oscillate too fast; that is, even if it is not necessarily a bounded-bandwidth function, their coefficients decay fast enough.
Such idea was applied to sleep apnea events detection \cite{lin2016sleep} and blood pressure analysis \cite{Wu2016modeling}. This model and its  generalizations have been considered to design different algorithms \cite{hou2016extracting,Yang2019multiresolution,Xu2018recursive}; for example, the case $\hat{s}(1)=0$ is discussed in \cite{Xu2018recursive}, but to simplify the discussion, we focus on the above assumption.

\subsection{Time-varying Wave-shape Functions and the proposed adaptive non-harmonic model}\label{Section:ANHM}

The model \eqref{eq:WSF} is however limited when it is applied to study more complicated signals. Specifically, in biomedical signals, the oscillatory patterns change from time to time, and a generalization is necessary.
To this end, note that by replacing $s(t)\in \mathcal{S}^{\delta,D,\theta}$ by its Fourier series in \eqref{eq:WSF}, the signal \eqref{eq:WSF} can be rewritten as
\begin{align}
f_{\texttt{fix}}&(t) = A(t) \sum_{\ell = -\infty}^{\infty} \hat{s}(\ell) e^{i 2\pi \ell \phi(t)}\label{eq:WSF_senos_cosenos}\\
    &= A(t) \sum_{\ell = 1}^{\infty} \left(\alpha_\ell \cos( 2\pi \ell\phi(t)) + \beta_\ell \sin( 2\pi \ell\phi(t))\right),\nonumber
\end{align}
where $\alpha_\ell = 2 \Re(\hat{s}(\ell))$, $\beta_\ell = -2\Im(\hat{s}(\ell))$, $\hat{s}(0) = 0$ by assumption, and $\Re$ and $\Im$ mean taking the real and imaginary parts. The form \eqref{eq:WSF_senos_cosenos} could be trivially further rewritten as
\begin{equation}\label{eq:WSF_cosenos}
f_{\texttt{fix}}(t) = A(t) \sum_{\ell = 1}^{\infty} a_\ell \cos( 2\pi \ell\phi(t) + b_\ell)\,,
\end{equation}
where $a_\ell = |\hat{s}(\ell)|/2$ and $b_\ell = \arctan(\Im(\hat{s}(\ell)) / \Re(\hat{s}(\ell)))$.
Physically, the model \eqref{eq:WSF} could be understood via the lens of \eqref{eq:WSF_cosenos}, that is, it is composed of possibly infinitely many oscillatory functions, where the $\ell$-th component is $[a_\ell A(t)]\cos( 2\pi \ell\phi(t) + b_\ell)$ with the AM $a_\ell A(t)$ and the phase $2\pi \ell\phi(t) + b_\ell$. However, when $\ell$ is large and $a_\ell> 0$, in general $[a_\ell A(t)]\cos( 2\pi \ell\phi(t) + b_\ell)$ is not an IMT function since the condition C2 may not hold on $\|\ell\phi''\|_\infty$.

To accommodate time-varying WSF, in this paper the model \eqref{eq:WSF_cosenos} was generalized to
\begin{equation}\label{eq:TVWSF2}
 f_{\texttt{var}}(t) := A(t)\sum_{\ell = 1}^\infty a_{\ell} \cos(2\pi \phi_{\ell}(t))\,,
\end{equation}
where $\{a_\ell\}_{\ell=1}^\infty\subset \mathbb{R}$ is an $\ell^1$ sequence, $A(t)$ and $\phi_{1}(t)$ satisfy the slowly varying condition C1 and C2, and $\phi_{\ell}(t)$ satisfy the condition C3:
\begin{itemize}
\item[C3.] $|\phi'_\ell(t) - \ell\phi'_1(t)| \leq \epsilon \phi'_1(t)$, for all $\ell = 1,\dots,\infty$. This condition ensures the IFs are not far from a multiple of the fundamental frequency.
\end{itemize}
The condition C3 says that the IF of multiples, $\phi'_\ell$ for $\ell>1$, are no longer necessarily multiples of the fundamental frequency $\phi'_1(t)$. As a result, we cannot use {\em one} WSF to describe the signal and the WSF is time-varying, and it cannot be represented by few sinusoidally oscillatory components due to the time-varying IF.
We abuse the notation and still call a function of the format \eqref{eq:TVWSF2} an IMT function, or simply a {\em mode}.

The ANHM we consider in this paper satisfies
\begin{equation}\label{eq:f}
\begin{aligned}
F_{\texttt{var}}(t) = \sum_{i = 1}^I  f_{\texttt{var},i}(t)\,,
\end{aligned}
\end{equation}
where $f_{\texttt{var},i}(t)=A_i(t) \sum_{\ell = 1}^\infty a_{i,\ell} \cos(2\pi \phi_{i,\ell}(t))$ satisfies \eqref{eq:TVWSF2} and we further assume that
\begin{itemize}
\item[C4.] there exists $d>0$ so that $\inf_t|\phi_{i,1}'(t)-\phi_{j,1}'(t)|\geq d$ when $i\neq j$;
\item[C5.] $\phi_{i,1}(t) \neq k \phi_{j,1}(t)$ for $k \in \mathbb{N}$ for any $i\neq j$.
\end{itemize}
Here, the condition C4 is about the separation of fundamental IFs ($\ell = 1$). Note that we do not put any separation condition on the multiple IFs. This is because when $i\neq j$, $\ell\phi_{i,1}'(t)$ and $\ell'\phi_{j,1}'(t)$ will be as close as possible for some $\ell,\ell'\geq 1$ and $\ell\neq \ell'$ by the Weyl's equidistribution theorem, and hence $\phi_{i,\ell}'(t)$ and $\phi_{j,\ell'}'(t)$. We mention that when $I>1$, if we want to keep physical interpretation, we shall define the phase of $F_{\texttt{var}}(t)$ to be a set of functions $\{\phi_{i,1}\}_{i=1}^I$ but not a single function. We refer readers with interest to  \cite{huang2020scientific} for further discussion.
The condition C5 is critical, as it guarantees the identifiability of the proposed multiple components model. Indeed, if  $\phi_{i,1}(t) = k \phi_{j,1}(t)$, the summation of $f_{\texttt{var},i}$ and $f_{\texttt{var},j}$ becomes $\tilde A_j(t) \sum_{\ell = 1}^\infty \tilde a_{j,\ell} \cos(2\pi \phi_{j,\ell}(t))$ for some $\tilde A_j(t)$ and $\{\tilde a_{j,\ell}\}$.
The mission is how to decompose each \emph{mode} $f_{\texttt{var},i}(t)$, with $i = 1,\dots,I$, from $F_{\texttt{var}}(t)$, which is possibly contaminated by an independent noise.

\begin{remark}
It is possible to consider a more general time-varying WSF model. For example, the model considered in \cite{CYLin2018} allows more freedom on the AMs, which reads
\begin{equation}\label{eq:TVWSF}
f_{\texttt{more}}(t) = \sum_{\ell = 1}^{\infty} B_\ell(t) \cos(2 \pi \phi_\ell(t))\,,
\end{equation}
where $B_1(t)$ and $\phi_{1}(t)$ satisfy the slowly varying condition C1 and C2, $\phi_\ell \in C^2{\mathbb(R)}$ satisfies C3, and $B_\ell \in C^1(\mathbb{R})$ further satisfies
$B_\ell(t) \leq c(\ell) B_1(t)$, for all $\ell = 1,\dots,\infty$, and with $c = \{c(\ell)\}_{\ell = 0}^{\infty}$ a non-negative $\ell^{1}$ sequence,
and
$|B'_\ell(t)| \leq \epsilon c(\ell) \phi'_1(t)$ and $|\phi''_\ell(t)|\leq \epsilon \ell \phi'_1(t)$, for all $\ell = 1,\dots,\infty$. This model is clearly more complicated than \eqref{eq:TVWSF2} since the AMs are further assumed to vary independently from each other. From the perspective of algorithm design for signal decomposition, the main difference between \eqref{eq:TVWSF} and \eqref{eq:TVWSF2} comes from the complication of parameters to be estimated. In \eqref{eq:TVWSF}, we need to estimate {\em many functions}, while in \eqref{eq:TVWSF2} we only need to estimate {\em fewer functions} and {\em some sequences}. To balance between the number of parameters to be estimated and the purpose of tracking time-varying WSF, we focus on \eqref{eq:TVWSF2}.
\end{remark}

\begin{remark}
The considered ANHM is in the real form since most practical signals are real. We can also consider the ANHM in the complex form; for example,
$ f^{\mathbb{C}}_{\texttt{var}}(t) := A(t)\sum_{\ell = 1}^\infty a_{\ell} e^{i2\pi \phi_{\ell}(t)}$,
where $A(t)$, $a_{\ell}$ and $\phi_\ell$ fulfill the same conditions as those for \eqref{eq:TVWSF2}. Clearly, \eqref{eq:TVWSF2} can be recovered from taking the real part of $ f^{\mathbb{C}}_{\texttt{var}}$, while recovering $ f^{\mathbb{C}}_{\texttt{var}}$ from \eqref{eq:TVWSF2} is in general challenging. See \cite{Nuttall1966} for example.
Note that handling signals in the complex form is in general easier, particularly in the low frequency region. Indeed, when analyzing $f_{\texttt{var}}$, the spectral leakage from the negative frequencies via the uncertainty principle is inevitable.
\end{remark}

\section{Algorithms}\label{sec:Algorithms}

The proposed algorithm is composed of two parts. The first part is the estimation of AM and phase, which has been well developed in the literature. Below, we detail this part for the sake of self-containedness. The second part is the novel optimization algorithm we propose in this work.
Below, denote the recorded signal as $y(t)=f(t)+\epsilon(t)$, where $f(t)$ will be assigned later and $\epsilon(t)$ is an independent stationary noise with zero mean and finite variance.

\subsection{Part 1: Estimation of amplitude and phase modulations}\label{sec:estimation}

\subsubsection{Recall SST and its variation}
For a properly behaved function $f$, like a bounded distribution,
the short-time Fourier transform (STFT) is defined as
\begin{equation}
V_f^g(t,\eta) = \int_{-\infty}^{+\infty} f(u) g(u-t) e^{-i 2 \pi \eta (u-t)} du\,,
\end{equation}
where $g(t)$ is the kernel function chosen by the user. Usually we consider a smooth and fast decaying function as the kernel, like a Gaussian function. The STFT offers a time-frequency (TF) representation of the signal \cite{Cohen1995}. If $f(t)$ is modeled by \eqref{eq:f}, then its STFT has a specific TF structure -- each sinusoidal oscillation, $A_i(t) a_{i,\ell} \cos(2\pi \phi_{i,\ell}(t))$, occupies a ``ribbon'' around its instantaneous frequency (IF) $\phi'_{i,\ell}(t)$, with $i = 1,\dots,I$ \cite{Carmona1999}. The width of those ribbons is determined by the support of the Fourier transform of $g$.

A common procedure to obtain more concentrated TF domains for each oscillation is SST \cite{Wu2011adaptive,Oberlin2014}. The method consists of locally estimate the IF as
\begin{equation*}
\tilde{\omega}_f(t,\eta) = \begin{cases}
\Re \left( \frac{1}{2\pi} \partial_t \arg(V_f^g(t,\eta)) \right), \text{ if } V_f^g(t,\eta) \neq 0\\
-\infty, \text{ otherwise}
\end{cases},
\end{equation*}
where $\arg$ is interpreted as a smooth argument function respecting its multi-valued nature,
and use it to vertically reassign (or synchrosqueeze) the STFT:
\begin{equation}
S_f(t,\omega) = \frac{1}{g(0)} \int V_f^g(t,\eta) \delta(\omega - \tilde{\omega}_f(t,\eta)) d\eta\, ,
\end{equation}

\noindent where $\delta(\cdot)$ is the Dirac delta measure. The synchrosqueezed STFT $S_f(t,\omega)$ offers a more concentrated TF representation, but it is accurate enough only for slowly varying IF \cite{Oberlin2015}. In order to improve the accuracy of SST, an estimator of the {\em group delay} is necessary:
\begin{equation}\label{eq:group_delay}
\tilde{\tau}_f(t,\eta) = \frac{1}{2\pi} \partial_\eta \arg(V_f^g(t,\eta)).
\end{equation}
Then, a new IF estimator can be built as
\begin{align*}
&\tilde{\omega}_f^{[2]}(t,\eta) \\
 = &\begin{cases}
\tilde{\omega}_f(t,\eta) + \Re\left( \tilde{q}_f(t,\eta)(t - \tilde{\tau}_f(t,\eta)) \right) \text{ if } \partial_t \tilde{\tau}_f(t,\eta) \neq 0\\
\tilde{\omega}_f(t,\eta) \text{ otherwise},
\end{cases}\nonumber
\end{align*}
 where $\tilde{q}_f(t,\eta) = \frac{\partial_t \tilde{\omega}_f(t,\eta)}{\partial_t \tilde{\tau}_f(t,\eta)}$, and whose real part constitutes an estimate of the so-called \emph{chirp rate} \cite{fourer2017chirp}. An efficient way to compute all the operators is by using five different STFTs $V_f^g, V_f^{tg}, V_f^{g'},V_f^{g''}$ and $V_f^{tg'}$, without the need for any differentiation \cite{Auger1995,Oberlin2015,Flandrin2018explorations}. Then, the \emph{second-order} SST (SST2) is obtained replacing $\tilde{\omega}_f$ by $\tilde{\omega}_f^{[2]}$ in standard SST \cite{Oberlin2015}:
\begin{equation}
S_f^{[2]}(t,\omega) = \frac{1}{g(0)} \int V_f^g(t,\eta) \delta(\omega - \tilde{\omega}_f^{[2]}(t,\eta)) d\eta.
\end{equation}
Compared with STFT, the TF representation provided by SST is sharper and more concentrated around IFs. With this feature, we can apply a ridge detection algorithm \cite{Meignen2017,Colominas2020} to extract the ridge associated with the first oscillatory component, denoted as $c_{11}(t)$. According to established theory \cite{Daubechies2011synchro}, we have $c_{1,1}(t) \approx \phi'_{1,1}(t)$; that is, $c_{1,1}$ constitutes an approximation to the IF $\phi'_{1,1}$.

An important feature of SST is that the TF representation remains invertible.
Therefore, a reconstruction of $A_1(t)a_{1,1} \cos(2\pi \phi_{1,1}(t))$ is possible via
\begin{equation}\label{eq:z}
\tilde{f}_1^{\mathbb{C}}(t) = \int_{|\omega - c_{1,1}(t)|<\Delta} S_f^{[2]}(t,\omega) d\omega\,,
\end{equation}
where $\Delta$ is a positive small constant chosen by the user. By \cite{Daubechies2011synchro}, we have $\tilde{f}_1^{\mathbb{C}}(t)\approx A_1(t)a_{1,1} e^{i2\pi \phi_{1,1}(t)}$.

The robustness of SST deserves some discussion, since the differentiation step might cause some alarms. As indicated above, the nature of SST does not depend on any differentiation \cite{Oberlin2015,Auger2013}. Theoretically, it has been shown that SST is robust to various kinds of noises, even non-stationary \cite{Chen_Cheng_Wu:2014}, and the asymptotic distribution of Gaussian random process has been established for the statistical inference purpose \cite{sourisseau2019inference}.

\subsubsection{Estimate $ A_i(t)$ and $\phi_{i,1}(t)$ from $y(t)$}
With the above described SST and its properties, we could now estimate $ A_i(t)$ and $\phi_{i,1}(t)$ from $y(t)$.
Indeed, the modulus and phase of the complex signal $\tilde{f}_1^{\mathbb{C}}(t)$ offer estimations of the AM and phase of the first oscillatory component:
\begin{equation}
\tilde{A}_1(t)\tilde{a}_{1,1} = |\tilde{f}_1^{\mathbb{C}}(t)|, \quad \tilde{\phi}_{1,1}(t) = \operatorname{phase}(\tilde{f}_1^{\mathbb{C}}(t)).
\end{equation}
The procedure continues with a {\em peeling scheme}. Denote $\tilde{S}_{y,1}^{[2]}:=\tilde{S}_y^{[2]}$. For each $j>1$ and the extracted ridge $c_{j,1}$, a new TF representation is defined as
\begin{equation}
\tilde{S}_{y,j+1}^{[2]}(t,\omega) = \begin{cases}
0, \text{ if } |\omega - k\, c_{j,1}(t)|<\Delta, \, k \in \mathbb{N} \\
S_{y,j}^{[2]}(t,\omega), \text{ otherwise}
\end{cases},
\end{equation}
and a new ridge is extracted from it (this time, $c_{j+1,1}(t) \approx \phi'_{j+1,1}(t)$), which leads to the estimates of ${A}_{j+1}(t){a}_{j+1,1}$ and $\phi_{j+1,1}(t)$. The procedure continues sequentially, until there is no more ridge to extract \cite{Meignen2017,ikram2001estimation}. As a result, we obtain an estimate of $I$, and all $I$ fundamental amplitudes and phases are estimated.

We should comment that SST is not the only possible choice for the phase and amplitude estimation. Any algorithms that can estimate the phase and amplitude accurately and robustly could be considered; for example, the empirical mode decomposition \cite{Huang1998}, the widely applied continuous wavelet transform \cite{Flandrin2018explorations} and the Blaschke decomposition \cite{Nahon:2000Thesis}.
See \cite{wu2020current} for a summary of various choices in the literature. We consider SST due to two reasons. First, it has been reported that the ridge detection performance is better compared with linear-type TF analysis tools when SST is applied \cite{Meignen2017} due to the sharper TF representation. Second, there is more freedom for the selection of the $\Delta$ parameter in \eqref{eq:z}, since the modes occupy a narrower space on the TF plane \cite{Wu2011adaptive,Oberlin2014}. See Fig. \ref{fig:06_STFT} for an example with details in Sec. \ref{sec:Results 1st simulated signal}.

\begin{remark}
There are several works mainly focusing on the IF estimation mission, for example, the improved sliding pairwise ICI rule approach \cite{lerga2011improved} and the multiview TF distributions based on the adaptive fractional spectrogram \cite{khan2012instantaneous}, among others \cite{lerga2011efficient}. However, we should comment that estimating the IF and estimating the phase are two related but different missions. It is possible to estimate the phase first and then obtain the IF by a direct differentiation. But estimating the phase from the estimated IF might not be an easy job due to the potential accumulated error from the numerical integration and the potential error from the initial phase estimation.
\end{remark}

\begin{remark}
We just described a procedure in which $\phi'_{i1}(t)$ are ``dominant'' in the TF representation; i.e. they are the most energetic ridges. When this is not the case, particularly when the fundamental frequency is not ``dominant'' or the corresponding ridge is not the one with the most energy, the IF estimation can be difficult when using a peeling scheme for ridge extraction \cite{Carmona1999,Colominas2020}. In this case, tools such as de-shape \cite{CYLin2018} could be applied, prior to ridge detection, to estimate the fundamental IF of each component. This is however out of the scope of this paper.
\end{remark}

\subsection{Part 2: Shape-adaptive mode decomposition (SAMD)}

We now introduce our proposed decomposition algorithm, shape-adaptive mode decomposition (SAMD), based on the the estimates $\tilde{A}_i(t) \approx A_i(t)$ and $\tilde{\Phi}_{i1}(t) \approx 2 \pi \phi_{i,1}(t)$, for $i = 1,\dots,I$. Consider the following optimization problem:
\begin{equation}
\begin{aligned}
\min_{\substack{c_{i\ell}, d_{i\ell} \\ \Phi_{i \ell}}}\, &\Big\| y(t) - \sum_{i = 1}^I \tilde{A}_i(t) \sum_{\ell = 1}^{D_i} \left(c_{i\ell} \cos({\Phi}_{i\ell}(t)) \right.  \\
&\quad +  \left. d_{i\ell}\sin({\Phi}_{i\ell}(t)) \right) \Big\|_2^2.
\end{aligned}
\end{equation}
where the phases $\Phi_{i \ell}$ must be fitted and $D_i$ should be estimated. We propose here to fit these phases as
\begin{equation}\label{eq:phi_model}
\Phi_{i \ell} = \sum_{k=1}^K e_{i \ell k} \tilde{\Phi}_{i1}^k, \, i = 1,\dots,I, \, \ell\geq2, \, k=1,\dots,K,
\end{equation}
where $e_{i11} = 1$ and $e_{i1k} = 0$ for $i = 1,\dots,I$ and $k \geq 2$, since we use the estimations as the first harmonic of each mode, i.e. $\Phi_{i1} = \tilde{\Phi}_{i1}$. We propose here to model the phases $\Phi_{i \ell}$ as linear combinations of the power $k$ of the estimations $\tilde{\Phi}_{i1}$, in a ``polynomial fitting'' fashion. This allows us to accommodate for more complex phases than those merely being integer multiples. Then, the evidently \emph{nonlinear regression} problem reads
\begin{equation}\label{eq:nonlinear_problem}
\begin{aligned}
\min_{\substack{c_{i\ell}, d_{i\ell} \\ e_{i \ell k}}} \, &\left\| y(t) - \sum_{i = 1}^I \tilde{A}_i(t) \sum_{\ell = 1}^{D_i} \left(c_{i\ell} \cos\left(\sum_{k=1}^K e_{i \ell k} \tilde{\Phi}_{i1}^k\right)\right. \right. \\
&\left.\left. \quad + d_{i\ell}\sin\left(\sum_{k=1}^K e_{i \ell k} \tilde{\Phi}_{i1}^k\right) \right) \right\|_2^2.
\end{aligned}
\end{equation}

The trigonometric functions, sine and cosine, applied to some of our regression coefficients (linear combinations involving $e_{i\ell k}$) make this problem nonlinear. In general, there is no closed-form expression for the optimal parameters (as opposed to linear regression in Sec. \ref{sec:linear_regression}), and iterative algorithms are needed to solve the problem. As with every iterative method, there is a need for initial values. We consider a warm-start strategy \cite{Yildirim2002}, using the solution of the linear regression approach (see Section \ref{sec:linear_regression} below) as initial guess for the coefficients $c_{i \ell}$ and $d_{i \ell}$. As for the coefficients $e_{i\ell k}$, we use the condition
\begin{equation}
|\tilde{\Phi}'_{i \ell}(t) - \ell \Phi'_{i1}(t)| \leq \epsilon \Phi'_{i1}(t)
\end{equation}
\noindent and set $e_{i \ell 1} = \ell$, and $e_{i \ell k} = 0$, for $i = 1,2$, $\ell = 1,\dots,D_i$, and $k = 2,\dots,K$, as initial values. In the following, we will call this method {\em shape-adaptive mode decomposition} (SAMD).

\subsection{Numerical implementation}\label{Section: numerical SAMD}

Suppose the signal $y(t)$ is discretized with the sampling frequency $f_s = 1/\Delta t$, with $\Delta t >0$ the sampling period, over the interval $[\Delta t, N \Delta t]$. Then, we end up with a (column) vector $\mathbf{y} \in \mathbb{R}^N$, with $\mathbf{y}(n) = y(n \Delta t)$ for $n = 1, \dots, N$, which is composed of the clean signal $\mathbf{f} \in \mathbb{R}^N$, with $\mathbf{f}(n) = F_{\texttt{var}}(n \Delta t)$ for $n = 1, \dots, N$ and the observational noise $\epsilon \Xi(n)$, where $\epsilon \in \mathbb{R}$ and $\mbox{var}(\Xi(n)) = 1$ for all $n$.
The estimated AMs and phases from Part 1 are denoted as $\tilde{\mathbf{A}}_i \in \mathbb{R}^N$, with $i = 1,\dots,I$ such that $\tilde{\mathbf{A}}_i(n)$ is an estimate of $A_i(n \Delta t)$ and $\tilde{\mathbf{\Phi}}_{i1} \in \mathbb{R}^N$, with $i = 1,\dots,I$ such that $\tilde{\mathbf{\Phi}}_{i1}(n)$ is an estimate of $2\pi \phi_{i,1}(n \Delta t)$. Then, \eqref{eq:nonlinear_problem} is discretized accordingly. We can construct the following matrix
\begin{equation}
\mathbf{Q}(\mathbf{e}) = [\mathbf{Q}_1(\mathbf{e}) \,\, \mathbf{Q}_2(\mathbf{e}) \, \, \cdots \mathbf{Q}_i(\mathbf{e})]^T \in \mathbb{R}^{(2\sum_{i = 1}^I D_i) \times N},
\end{equation}
\noindent with $\mathbf{e}=\{e_{i,\ell,k}\}$,
$\mathbf{Q}_i(\mathbf{e}) = [\mathbf{p}_{i1},\dots,\mathbf{p}_{iD_i},\mathbf{q}_{i1},\mathbf{q}_{iD_i}]^T \in \mathbb{R}^{2 D_i \times N}$,
and the $N$-dim vectors
\begin{align}
\mathbf{p}_{i\ell}(n) &:=  \tilde{\mathbf{A}}_i(n) \cos\left( \sum_{k=1}^K e_{i,\ell,k}\tilde{\mathbf{\Phi}}^k_{i1}(n)\right), \nonumber\\
 \mathbf{q}_{i\ell}(n) &:=  \tilde{\mathbf{A}}_{i}(n) \sin\left( \sum_{k=1}^K e_{i,\ell,k} \tilde{\mathbf{\Phi}}^k_{i1}(n)\right),\nonumber
\end{align}
\noindent for $n = 1,\dots,N$.
Then, our recorded observation can thus be written as
\begin{equation}
\mathbf{y} = \mathbf{c}^T \mathbf{Q}(\mathbf{e}) + \epsilon\mathbf{\Xi},
\end{equation}
\noindent where $\mathbf{c} = [\mathbf{c}_1^T \, \mathbf{c}_2^T \, \cdots \mathbf{c}_i^T]^T \in \mathbb{R}^{2\sum_{i = 1}^I D_i}$, with $ \mathbf{c}_i = [c_{i1},\dots,c_{iD_i},d_{i1},\dots,d_{iD_i}]^T \in \mathbb{R}^{2D_i}$, for $i = 1,\dots,I$.

Then, estimate $\mathbf{c}$ and $\mathbf{e}$ by solving the following optimization problem (remember that $\mathbb{E}(\Delta t \mathbf{\Xi} \mathbf{Q(e)}^T) = 0$):
\begin{equation}\label{Nonlinear Optimization Problem discretization}
\min_{\mathbf{c},\mathbf{e}} \| \mathbf{y} - \mathbf{c}^T \mathbf{Q}(\mathbf{e}) \|_2^2,
\end{equation}
where $\| \cdot \|_2$ stands for the $\ell_2$ norm. Denote the minimizers as $\tilde{\mathbf{c}}$ and $\tilde{\mathbf{e}}$. We solve our nonlinear problem \eqref{Nonlinear Optimization Problem discretization} with MATLAB's \verb"nlinfit" function. We use the Cauchy weight function for robust fitting: \verb"options.RobustWgtFun = 'cauchy'". In this case, the function uses an iterative reweighted least squares algorithm \cite{dumouchel1989integrating,holland1977robust}. At each iteration, the robust weights are recomputed according to the residual of each observation (discrete sample) from the previous iteration. In this way, the influence of the outliers on the fit is decreased. Finally, a decomposition might be achieved by
\begin{equation}\label{Formula: reconstruction by SAMD}
\tilde{f}_{\texttt{var},i}(n \Delta t) = \tilde{\mathbf{c}}_i^T \mathbf{Q}_i(\tilde{\mathbf{e}}),
\end{equation}
where $i = 1,\dots,I$, and the clean signal $f(t)$ can be approximated as
$\tilde{F}_{\texttt{var}}(n \Delta t) = \tilde{\mathbf{c}}^T \mathbf{Q}(\tilde{\mathbf{e}})$.

The pseudo-code for SAMD can be found in Algo. 1. The discretization and implementation parameters for Part 1 and Part 2 will be detailed in the next section, and for the reproducibility purpose,
our codes can be found in \url{https://github.com/macolominas/SAMD}.

\hrulefill

\textbf{Algorithm 1} Shape-adaptive mode decomposition (SAMD)
\vspace{-6mm}

\hrulefill

\begin{algorithmic}[1]
\STATE \textbf{Input:} signal $y(t)$, and $K$ (for phases estimations).
\STATE Estimate the amplitudes $\tilde{A}_i(t)$ and phases $\tilde{\phi}_{i1}(t)$ from second-order SST, ridge detection, and partial reconstruction (Sec. \ref{sec:estimation}), which also give an estimate of $I$.
\STATE Solve the linear regression problem to obtain the coefficients $\hat{c}_{i,\ell}$ and $\hat{d}_{i,\ell}$,  which also give an estimate of parameters $D_i$ (Sec. \ref{sec:linear_regression}).
\STATE Solve the nonlinear regression problem from Eq. \eqref{eq:nonlinear_problem} using $\hat{c}_{i,\ell}$ and $\hat{d}_{i,\ell}$, and $e_{i\ell 1} = \ell$, $e_{i \ell k} = 0$ for $k = 2,\dots,K$, as initial values.
\STATE With the coefficients $\tilde{c}_{i,\ell}$, $\tilde{d}_{i,\ell}$ and $\tilde{e}_{i \ell k}$ synthesize the modes $f_{\texttt{var},i}(t)$.
\STATE \textbf{Output:} modes $f_{\texttt{var},i}(t)$, $i = 1,\dots,I$.
\end{algorithmic}

\vspace{-3mm}
\hrulefill

\subsection{Existing algorithms}
\subsubsection{Linear regression (LR)}\label{sec:linear_regression}

We summarize the LR algorithm designed to handle \eqref{eq:WSF} with fixed WSFs \cite{Wu2016modeling}. Suppose $y(t) = \sum_{i = 1}^I f_{\texttt{fix},i}(t) + \epsilon(t)$, where $f_{\texttt{fix},i}(t)=A_i(t) s_i(\phi_i(t))$ and $A_i(t) s_i(\phi_i(t))$ satisfies \eqref{eq:WSF} and $\epsilon(t)$ is an independent stationary noise with zero mean and finite variance. By \eqref{eq:WSF_senos_cosenos} and \eqref{eq:WSF_cosenos}, we have
\begin{equation}\label{eq:fixed}
y(t) = \sum_{i = 1}^I \sum_{\ell = 1}^{D_i} A_i(t) a_{i,\ell} \cos( 2\pi \ell\phi_i(t) + b_{i,\ell}) + \epsilon(t),
\end{equation}
where we approximated $s_i(t)$ by its first $D_i$ harmonics. Define $\mathbf{D} := \{D_1,\dots,D_I\}$ to be the set of parameters $D_i$. Given estimates $\tilde{A}_i(t) \approx A_i(t)$ and $\tilde{\Phi}_i(t) \approx 2 \pi \phi_{i}(t) + b_{i,1}$, for $i = 1, \dots,I$, a decomposition of the signal can be obtained. The problem can be formulated in terms of sines and cosines, with coefficients $c_{i \ell}$ and $d_{i\ell}$, for $i=1,\dots,I$, and $\ell$ from $1$ to $D_i$, where it is enough for the coefficients to be real:
\begin{equation}\label{eq:problem}
\begin{aligned}
\min_{c_{i\ell}, d_{i\ell}} \, &\left\| y(t) - \sum_{i = 1}^I \tilde{A}_i(t) \sum_{\ell = 1}^{D_i} \left(c_{i\ell} \cos(\ell \tilde{\Phi}_i(t)) \right. \right. \\
&\quad + \left. \left. d_{i\ell}\sin(\ell \tilde{\Phi}_i(t)) \right) \right\|_2^2.
\end{aligned}
\end{equation}
With the same discretization scheme of Sec. \ref{Section: numerical SAMD},
following \cite{Wu2016modeling}, we construct a matrix that plays a similar role to $\mathbf{Q}(\mathbf{e})$,
\begin{equation}
\mathbf{S} = [\mathbf{S}_1 \,\, \mathbf{S}_2 \, \, \cdots \mathbf{S}_i]^T \in \mathbb{R}^{(2\sum_{i = 1}^I D_i) \times N},
\end{equation}
\noindent with
$\mathbf{S}_i = [\mathbf{r}_{i1},\dots,\mathbf{r}_{iD_i},\mathbf{s}_{i1},\mathbf{s}_{iD_i}]^T \in \mathbb{R}^{2 D_i \times N}$,
and the $N$-dim vectors
\begin{equation}
\mathbf{r}_{i\ell}(n) :=  \tilde{\mathbf{A}}_i(n) \cos( \ell \tilde{\mathbf{\Phi}}_i(n)), \, \, \mathbf{s}_{i\ell}(n) :=  \tilde{\mathbf{A}}_i(n) \sin( \ell \tilde{\mathbf{\Phi}}_i(n)),\nonumber
\end{equation}
\noindent for $n = 1,\dots,N$, and solve the \emph{linear regression} problem
$\min_{\mathbf{c}} \| \mathbf{y} - \mathbf{c}^T \mathbf{S} \|_2^2$, where $\mathbf{c}$ is defined as before and $\mathbf{y}$ is the discretization of $y(t)$. This problem can be solved precisely by $\hat{\mathbf{c}} := (\mathbf{y} \mathbf{S}^T)(\mathbf{S}\mathbf{S}^T)^{-1}$, from which the decomposition might be achieved by $\tilde{f}^{\mathbf{D}}_{\texttt{fix},i} = \hat{\mathbf{c}}_i^T \mathbf{S}_i$, where $i = 1,\dots,I$, and denote $\tilde{F}^{\mathbf{D}}_{\texttt{fix}} = \hat{\mathbf{c}}^T \mathbf{S}$ as an estimation of the clean signal.

\subsubsection{Recursive diffeomorphism-based regression}

A different method that looks for a solution to the model \eqref{eq:fixed} is the so-called recursive diffeomorphism-based regression (RDBR) \cite{Xu2018recursive}. For each $i=1,\ldots, I$, denote the estimators of $A_i(t)$ and $\phi_i(t)$ as $\tilde A_i(t)$ and $\tilde \phi_i(t)$, and  set $r^{(0)}=y$. For each iteration $j=0,1,2,\ldots, J-1$, where $J\in \mathbb{N}$ is determined by the user, the demodulated (unwrapped) $r^{(j)}$ is calculated
\begin{equation}
h^{(j)}_i(t) = \frac{r^{(j)}(\tilde \phi_i^{-1}(t))}{\tilde A_i( \tilde\phi_i^{-1}(t))},
\end{equation}
which, ideally, constitute constant-amplitude $1$-periodic functions. The folding map $(t, h^{(j)}_i(t)) \mapsto (\operatorname{mod}(t,1),h^{(j)}_i(t))$ allows rough estimations of the fixed WSFs $\tilde{s}^{(j)}_i(t)$ through functional regression. Then the method continues in a deflationary manner on $r^{(j+1)}:=r^{(j)} - \sum_i\tilde A_i(t) \tilde{s}^{(j)}_i(\tilde \phi_i(t))$ to refine the WSFs estimations.

\subsubsection{Multiresolution mode decomposition}

The multiresolution mode decomposition (MMD) \cite{Yang2019multiresolution}, is an effort to generalize RDBR.
In the particular case of model \eqref{eq:f}, MMD would look for a decomposition of the form $f(t) = \sum_{i}^I f_i(t)$, where
\begin{equation}
\begin{aligned}
f_i(t) &= \sum_{n = -N/2}^{N/2} a_{n,i} \cos(2 \pi n \phi(t)) s_{cn,i}(2\pi N_i \phi(t)) \\
 &\quad + \sum_{n = -N/2}^{N/2} b_{n,i} \sin(2 \pi n \phi(t)) s_{sn,i}(2\pi N_i \phi(t)),
\end{aligned}
\end{equation}
 for $i = 1,\dots,I$, where $a_{n,i},b_{n,i}\in \mathbb{R}$ are coefficients, and $s_{cn,i}$ and $s_{sn,i}$ are real value functions modeling nonlinear and non-stationary data with time-dependent amplitudes, frequencies, and WSF. Note that this model is closer to the model \eqref{eq:TVWSF}. MMD applies the same deflationary algorithm as RDBR to estimate $a_{n,i}$, $b_{n,i}$, $s_{cn,i}$ and $s_{sn,i}$, but to only phase-demodulated versions of the signal (as before, using estimations of the phases $\tilde\phi_i(t)$) multiplied by a particular function, so we omit details and refer readers to \cite{Yang2019multiresolution}. Roughly speaking, from $f(\tilde \phi_i^{-1}(t))\cos(2 \pi m \tilde \phi_i(t))$, the method estimates $a_{m,i}$ and $s_{cn,i}$, and from $f(\tilde \phi_i^{-1}(t))\sin(2 \pi m \tilde \phi_i(t))$, it estimates $b_{m,i}$ and $s_{sn,i}$.

\subsection{Comparison of existing algorithms and our proposal}

Even though the four described algorithms need estimations of the amplitudes $A_i(t)$ and phases $\phi_{i,1}(t)$, there are important differences that must be remarked. The first point is regarding the model they aim to solve. For LR and RDBR, they both try to solve the model \eqref{eq:fixed}, i.e. a model with fixed WSF. While RDBR proposes an iterative scheme demodulating the signal and using some functional regression (which must be prescribed \emph{a priori}), LR uses a direct approach to estimate the Fourier coefficients of the WSFs. This results in a significant computational load for RDBR when compared to LR.

The methods of SAMD and MMD try, instead, to solve the more general model \eqref{eq:f}, i.e. a model which allows for time-varying WSFs. To do this, MMD proposes a generalization of RDBR, estimating functions $s_{cn,i}$ and $s_{sn,i}$ that describe the time-varying WSF. As with RDBR, an iterative deflationary scheme is needed, along with a functional regression proposal. A significant amount of time is needed, and for some signals the method might not converge.

Our SAMD proposal is a direct generalization of LR. Using the output of LR as initial values, along with a more general model for the phase functions (they no longer need to be an integer multiple of a fundamental phase), a flexible tool is achieved. We will show that our proposal is able to estimate modes with time-varying WSFs with good accuracy, and in a reasonable time.

Last but not the least, we mention that in the particular case where $I = 1$, the \emph{decomposition} mission is reduced to a \emph{denoising} problem. In the next section, the performance of the proposed SAMD and other algorithms for this purpose will also be demonstrated.

\section{Numerical Results}\label{sec:Results}

\subsection{A database of simulated signals}\label{sec:Results 1st simulated signal}

Let us consider $F(t) = \mathsf s_1(t) +  \mathsf s_2(t)$, with
\begin{equation}\label{eq:s1_6}
\begin{aligned}
 \mathsf s&_1(t) = 1.5 \cos(\phi_{11}(t)) + 0.25\cos((2.05+\xi_{2}) \phi_{11}(t)) \\
&+ \sum_{p = 3}^{10} 0.1 \cos((p + 0.05 + \xi_p) \phi_{11}(t) + 0.01\phi_{11}^2(t)),
\end{aligned}
\end{equation}
where $\phi_{11}(t) = 2\pi 6 t + 2\pi 6 t^2 + Y(t)$, $Y(t)$ is a random process and $\xi_p \in \mathbb{R}$ will be specified later, and
\begin{equation}\label{eq:s2_6}
 \mathsf s_2(t) = \cos(\phi_{21}(t)) \!+\! \sum_{p = 2}^{10} \frac{\cos((p+0.01+\zeta_p)\phi_{21}(t))}{\sqrt{p}},
\end{equation}
where $\phi_{21}(t) = 2\pi 10 t + 2\pi 7 t^2 + 0.5 \cos(2 \pi t) + Y(t) + Z(t)$, $Z(t)$ is a different random process and $\zeta_p\in \mathbb{R}$ will be specified later.
The signals are defined on $t \in [0,1]$ and sampled at 1000 Hz. Clearly, when $\xi_p=\zeta_p=0$ for $p=2,\dots,10$ and $Y(t)=Z(t)=0$, $F$ satisfies the ANHM and the condition C4 is satisfied. Also, neither $\mathsf s_1(t)$ nor $\mathsf s_2(t)$ can be written as a single waveform modulated by the phase function, and they cannot be written as a sum of a few sinusoidal oscillatory components.
Moreover, it is evident that $\phi'_{13}(t) = 3.05\phi'_{11}(t)$ and $\phi'_{22}(t) = 2.01\phi'_{21}(t)$ cross each other. The same happens for $\phi'_{14}(t)$ and $\phi'_{23}(t)$.
The changes on the waveforms can be appreciated on the second row of Fig. \ref{fig:06_ruido}. The situation is more complicated when $\xi_p$, $\zeta_p$, $Y(t)$ and $Z(t)$ are not zero. Such complicated signals constitute a good example to compare different algorithms.

We evaluate the performance of four methods: LR, SAMD, RDBR and MMD. For RDBR and MMD, we use the codes available at https://github.com/HaizhaoYang.
The amplitudes and phases are estimated  with the following setup (see Sec. \ref{sec:estimation}).
For the STFTs, we use a Gaussian window $g(t) = \sigma e^{-\frac{\pi t^2}{\sigma^2}}$, with $\sigma = 0.25$ (which minimizes the criterion of the R\'enyi entropy \cite{Baraniuk2001,Meignen2020}). We apply the second-order SST, and detect the ridges with the algorithm from \cite{Meignen2017} (allowing a maximum jump of 2 Hz between consecutive time instants). For the estimation of the complex function $\tilde{f}_1^{\mathbb{C}}(t)$ \eqref{eq:z}, we use $\Delta = 0.5$ Hz.
Regarding parameters $D_i$, we estimate them by adapting trigonometric regression tools, where we minimize a criterion which is a function of the model order, looking for a trade-off between error and model order \cite{Kavalieris1994,Ruiz2020}. Specifically, we can construct a criterion of the form $ \Omega(\mathbf{D}) = \|y - \tilde{F}_{\texttt{fix}}^{\mathbf{D}} \|_2 + G(\sum_i D_i)$, where $G(\sum_i D_i)$ is an increasing functional penalizing the model size. Then, the solution of $\min_{\mathbf{D}} \Omega(\mathbf{D})$ is the set of parameters to be used. Promising results of applications of these criteria on the WSF model can be found in \cite{Ruiz2020}.

\begin{figure}[t]
\begin{center}
\includegraphics[width=\columnwidth]{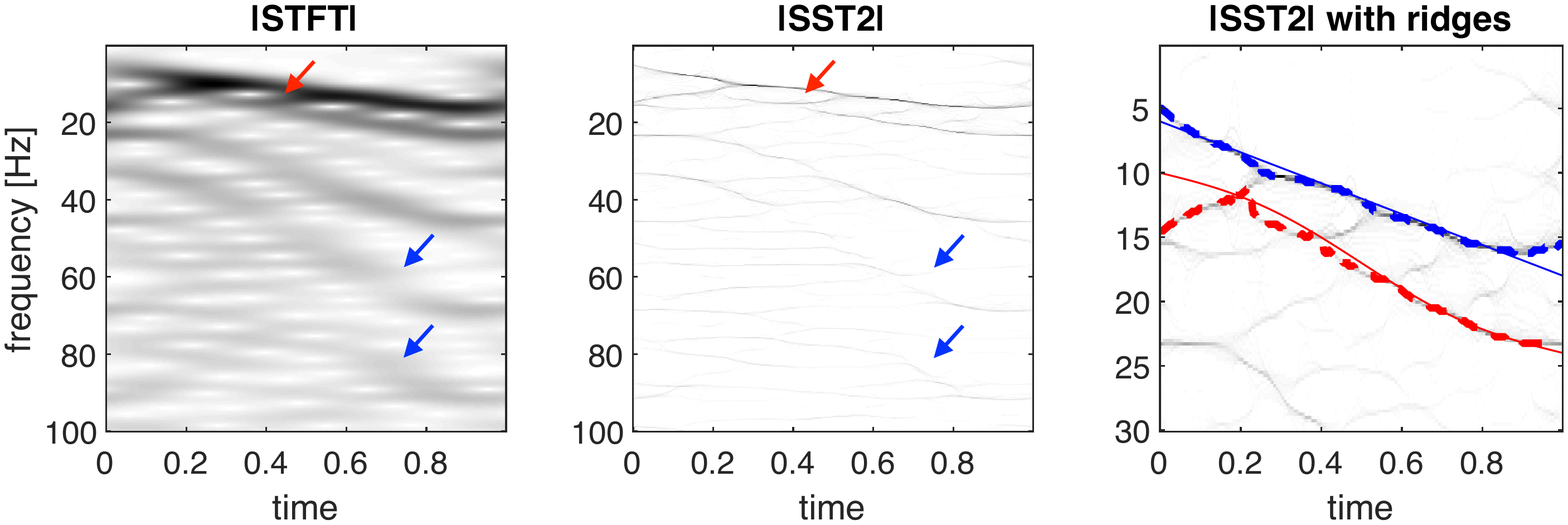}
\end{center}
\caption{\textbf{STFT and SST2 from noisy version of the sum of \eqref{eq:s1_6} and \eqref{eq:s2_6}.} Left: modulus of the STFT. Middle: modulus of SST2. Right: modulus of SST2 with the ground truth and detected ridges superimposed.}
\label{fig:06_STFT}
\end{figure}

An example of the TF representations of a noisy $F(t)$ (Gaussian white noise at 10 dB level) determined by STFT and SST2  can be appreciated in Fig. \ref{fig:06_STFT}, where we superimpose the ground truth as thin lines and detected ridges as thick dashed lines. The enhancement of the TF representations, and hence the separation of the two modes, can be found around the fundamental IFs that are indicated by red arrows. Due to the sharpness of both modes by SST2, the ridges associated with their multiples are less dominant as is indicated by blue arrows.

\subsubsection{Fixed $\xi_j$, $\zeta_p$, $Y(t)$ and $Z(t)$}
In this case, $\xi_p=\zeta_p=0$ for $p=2,\ldots,10$ and $Y(t)=Z(t)=0$. We used $K = 3$ for the estimations of the phases for SAMD as in Eq. \eqref{eq:phi_model}.
$D_1 = D_2 = 10$ were obtained for both LR and SAMD.

\begin{table}\label{table:03}
\caption{Errors and computation times for simulated signal from Eqs. \eqref{eq:s1_6} and \eqref{eq:s2_6} with $\xi_p=\zeta_p=0$ for $p=2,\dots,10$. *Out of the 100 realizations, MMD converged on 96 occasions.}
\begin{center}
\begin{tabular}{lcccc}
\hline
   & Noiseless & \multicolumn{3}{c}{Noisy (10 dB; 100 realizations)} \\
             & RMSE   & mean(RMSE) & std(RMSE)& mean time (s) \\
\hline
$\mathsf s_1$ (SAMD) & 0.206 &  0.288  & 0.017 &\multirow{2}{*}{43.210} \\
$\mathsf s_2$ (SAMD) & 0.501 & 0.542 & 0.039 &  \\
%$s_3$ (SAMD) & 0.1416 & 0.1740 & 0.0056 &  \\
\hline
$\mathsf s_1$ (LR) &  0.328  &  0.349   & 0.019 &\multirow{2}{*}{0.003}\\
$\mathsf s_2$ (LR) & 0.627   & 0.638    & 0.043 &  \\
%$s_3$ (LR) & 0.1438   & 0.1650    & 0.0047 &  \\
\hline
$\mathsf s_1$ (RDBR) & 0.330  &  0.355  & 0.021 &\multirow{2}{*}{14.245} \\
$\mathsf s_2$ (RDBR) & 0.631 &  0.638  & 0.042 &  \\
%$s_3$ (RDBR) & 0.2121  &  0.1635  & 0.0165 &  \\
\hline
$\mathsf s_1$ (MMD*) & 0.483 &  0.646 & 0.473 &\multirow{2}{*}{462.786}  \\
$\mathsf s_2$ (MMD*) & 0.483 &  0.611 & 0.063 &  \\
%$s_3$ (MMD*) & 0.1480 &  0.3411 & 0.0137 &  \\
\end{tabular}
\end{center}
\end{table}

\begin{figure}[t]
\begin{center}
\includegraphics[width=\columnwidth]{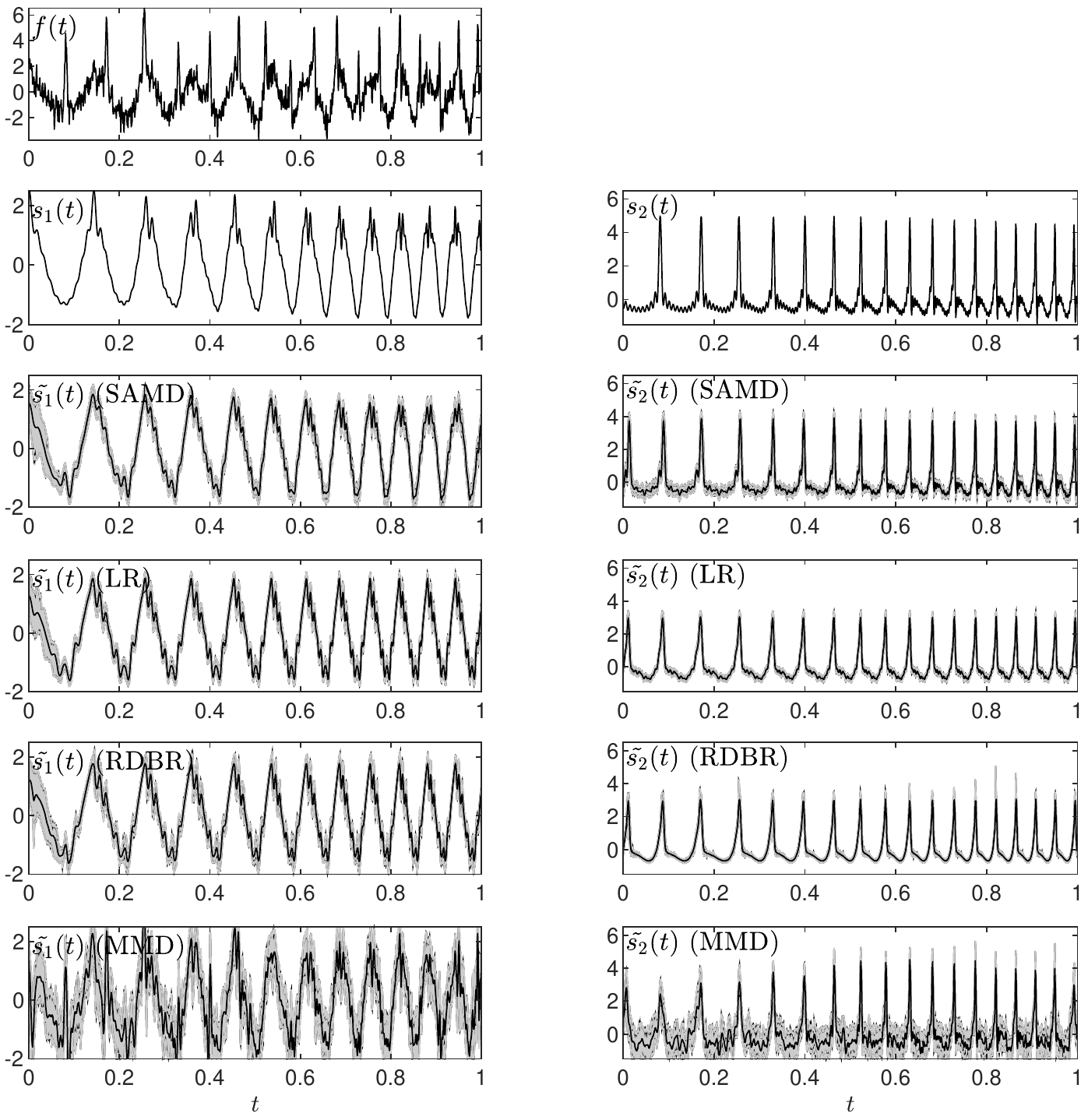}
\end{center}
\caption{\textbf{Simulated signal (noisy signals at 10 dB; 100 realizations).} First row: a typical noisy example of signal $F(t)=\mathsf s_1(t)+\mathsf s_2(t)+\epsilon(t)$ from Eqs. \eqref{eq:s1_6} and \eqref{eq:s2_6} with $\xi_p=\zeta_p=0$ for $p=2,\dots,10$. Second row: true components. Third row: extracted modes with our proposal SAMD. Fourth row: extracted components with LR. Fifth row: extracted components with RDBR. Sixth row: extracted components with MMD. (black solid line: mean estimated components; shaded gray area: 95\% confidence interval).}
\label{fig:06_ruido}
\end{figure}

In order to compare the robustness of different methods, we realized a noisy version of the signal at 10 dB for 100 times. The noise is assumed to be Gaussian white. We present these results on Fig. \ref{fig:06_ruido}. For each mode, we show the mean and the 95\% confidence interval of the root mean squared errors (RMSE). The mean and standard deviation of the RMSEs can be found at Table I, along with the averaged computational time. We see that SAMD has not only a better mode recovery performance, but also a comparable computational load when compared to RDBR, and a significantly lower burden compared to MMD (at least one order of magnitude).

\begin{figure}[t]
\begin{center}
\includegraphics[width=\columnwidth]{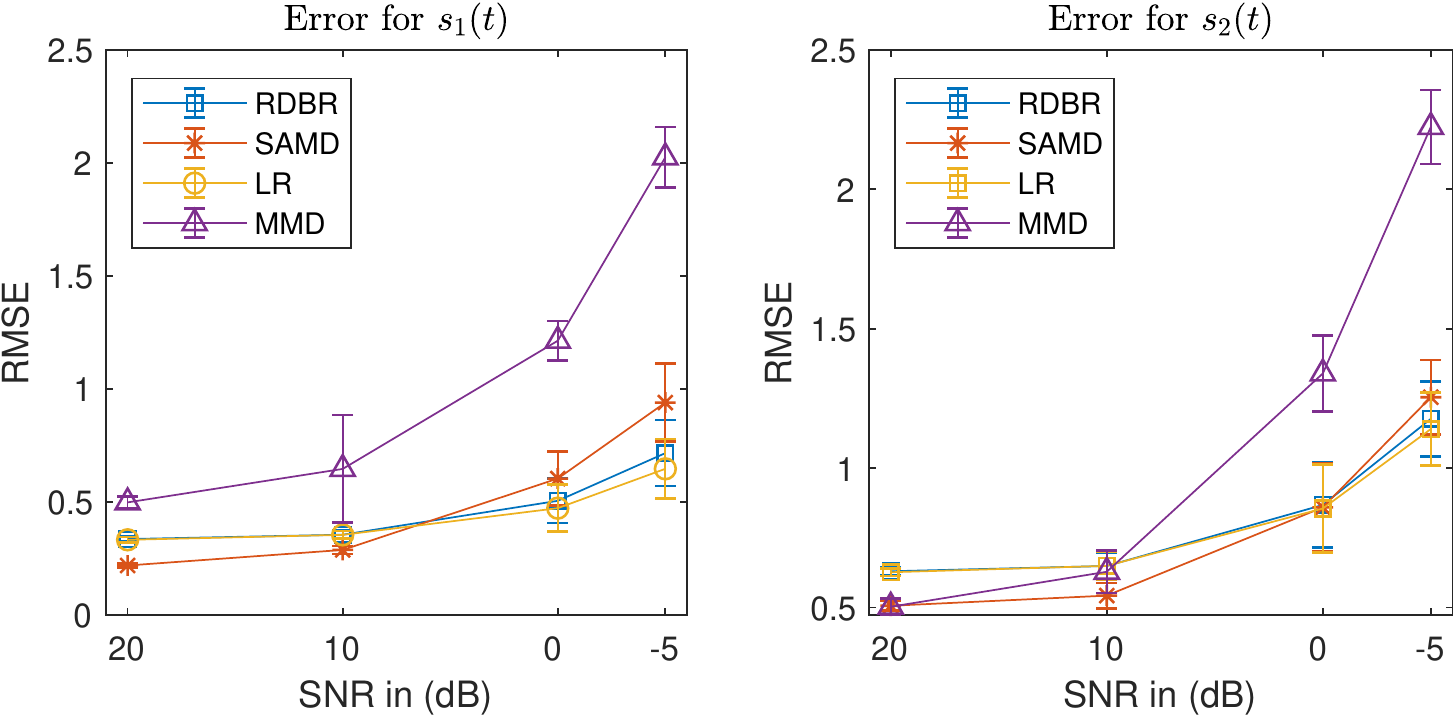}
\end{center}
\caption{\textbf{Errors for the simulated signal with $\xi_p=\zeta_p=0$ for $p=2,\dots,10$ (different SNRs; 50 realizations).} Left: mean errors and standard deviations for $\mathsf s_1(t)$. Right: mean errors and standard deviations for $\mathsf s_2(t)$. For the MMD method, we considered only those decompositions that converged (between 45 and 48 times out of the 50 realizations, depending on the input SNR).}
\label{fig:06_varias}
\end{figure}

We also tested robustness at different SNRs. We performed 50 decompositions of noisy versions of the signal at 20, 10, 0, and -5 dB, and computed the RMSEs for $\mathsf s_1(t)$ and $\mathsf s_2(t)$. Results can be appreciated on Fig. \ref{fig:06_varias}, where we present the mean and standard deviation of RMSEs over the 50 decompositions.
For $\mathsf s_1(t)$, the results of SAMD are the best for 20 and 10 dB and slightly worse than LR and RDBR for 0 and -5 dB.
For $\mathsf s_2(t)$, the results of SAMD are comparable to those of MMD for 20 dB, the best for 10 dB, comparable to those of LR and RDBR for 0 dB, and slightly worse than LR and RDBR for -5 dB. Out of the four analyzed methods, MMD seems to be by far the most sensitive to noise (and it did not converge for all 50 realizations).

\begin{figure}[t]
\begin{center}
\includegraphics[width=\columnwidth]{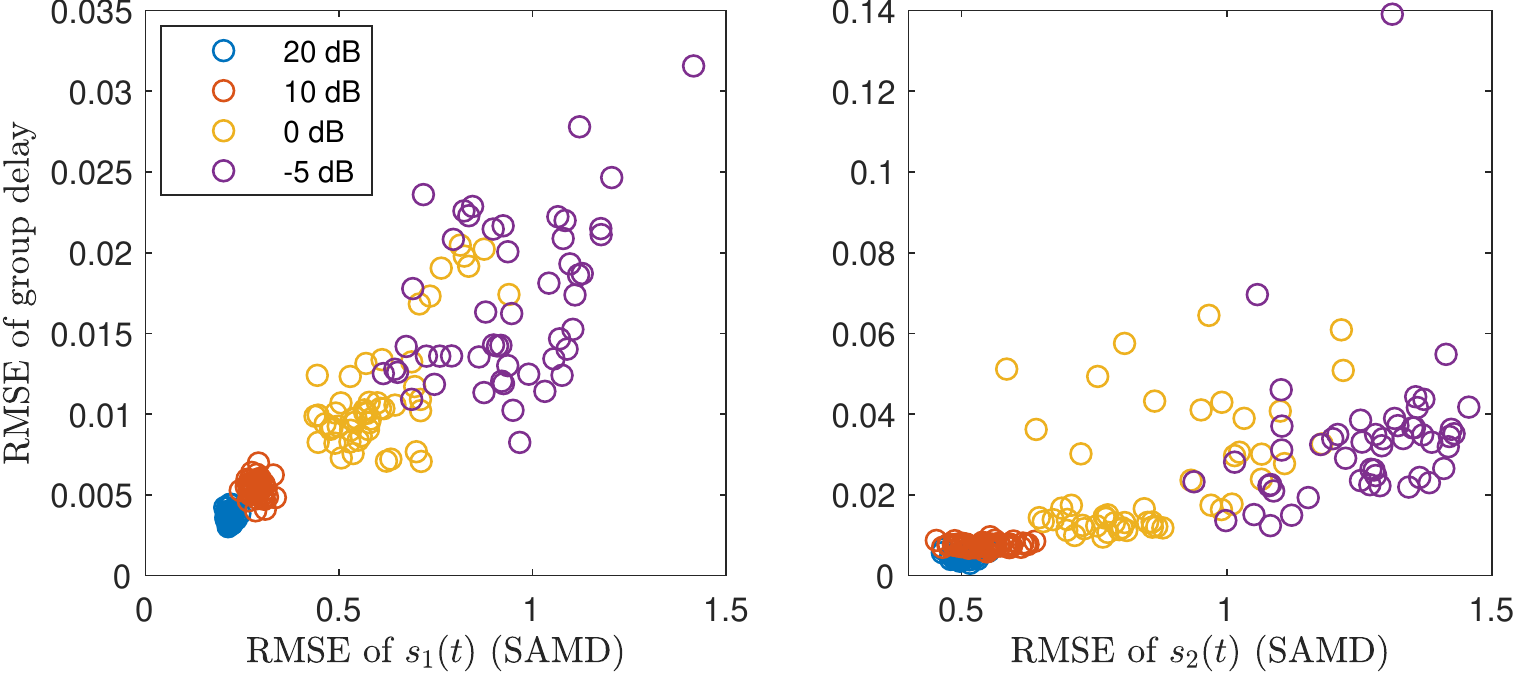}
\end{center}
\caption{\textbf{Errors for the group delay with $\xi_p=\zeta_p=0$ for $p=2,\dots,10$ (different SNRs; 50 realizations).} Left: RMSE of the group delay against RMSE of $\mathsf s_1(t)$. Right: RMSE of the group delay against RMSE of $\mathsf s_1(t)$.}
\label{fig:06_tau}
\end{figure}

The evaluation of the group delay estimation \eqref{eq:group_delay} necessary for SST2 is shown in Fig. \ref{fig:06_tau}. We evaluated this two-variable complex function ($\mathbb{R}^2 \mapsto \mathbb{C}$) on the detected ridge, and computed the error against the ideal group delay on the theoretical ridge (i.e. on the theoretical IF): $\|\tilde{\tau}_F(t,\phi'_{i,1}(t)) - \tilde{\tau}_{F+\epsilon}(t,c_{i,1}(t))\|_2$, where $F$ is the clean signal, and $\epsilon$ is the independent Gaussian white noise. As expected, the RMSE of the group delay is linearly correlated with the RMSE of $\mathsf s_1(t)$. The departure from the linear relation observed for $\mathsf s_2(t)$ in some realizations is due to the errors on the ridge detection, which is expected since the extraction of the first ridge might create some residues.

\subsubsection{Random $\xi_j$, $\zeta_p$, $Y(t)$ and $Z(t)$}

In this case, first consider the construction of $Y$ and $Z$. Let $W(t)$ be the standard Brownian motion, and $X_B(t) = W\ast K_B(t)$ its smoothed version, where $K_B(t)$ is the Gaussian function with standard deviation $B>0$ and $\ast$ denotes the convolution operator. Then, the random process
\begin{equation}
R_B(t) = 2\pi \int_{0}^t \frac{X_B(u)}{\|X_B\|_{L^\infty[0,1]}}  du,
\end{equation}
is defined for $t\in [0,1]$. Set $Y(t) = R_{80}(t)$ and $Z(t) = R_{50}(t)$. We assume that $Y(t)$ and $Z(t)$ are independent. Next, consider $\xi_2,\dots,\xi_{10} \sim \mathcal{U}[0.05,0.1]$, and $\zeta_3,\dots,\zeta_{10}\sim \mathcal{U}[0.01,0.02]$, where we assume that $\xi_j$, $\zeta_p$, $Y(t)$ and $Z(t)$ are independent.

Now, we decomposed $y(t) = F(t) + \epsilon(t)$, where $\epsilon(t)$ could be either Gaussian white noise, ARMA(1,1) noise, or Poisson noise (with a relative amplitude of 10 dB).  For the ARMA(1,1) case, the autoregressive and moving-averaging polynomials are chosen to be $0.5z + 1$ and $-0.5z + 1$ respectively with the i.i.d. Student $t_4$ random variables as the innovation process. Due to the `fat-tail' of Student $t_4$, the noise might be spiky.

\begin{figure}[t]
\begin{center}
\includegraphics[width=\columnwidth]{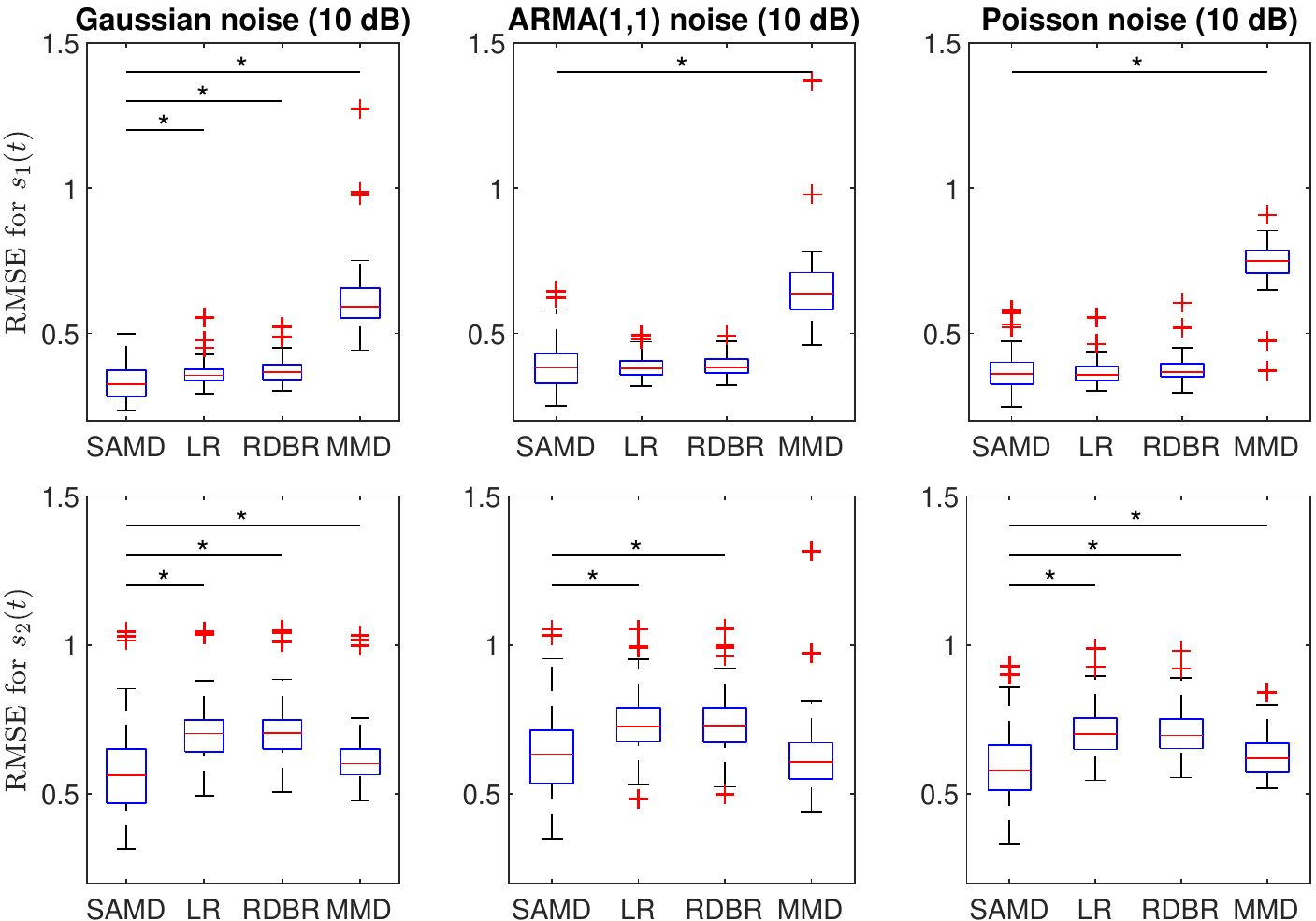}
\end{center}
\caption{\textbf{Results for simulated database when $\xi_2,\dots,\xi_{10} \sim \mathcal{U}[0.05,0.1]$, and $\zeta_2,\dots,\zeta_{10}\sim \mathcal{U}[0.01,0.02]$, and disturbed phases (10 dB; 100 realizations).}
Left: results for Gaussian white noise. Middle: results for ARMA(1,1) noise. Right: results for Poisson noise ($\lambda = 1$). `*': test in favor of SAMD. MMD converged between 95 and 96 times out of the 100 realizations.}
\label{fig:06_aleatorizando}
\end{figure}

The boxplot results for RMSE are presented in Fig. \ref{fig:06_aleatorizando}. MMD converged between 95 and 96 times out of the 100 realizations (depending on the type of noise), and RDBR diverged on one occasion. It can be appreciated that SAMD is the best among the four methods in the sense of the median. Wilcoxon signed rank tests (5\% significance level) were further carried out to compare different methods. Out of the six cases (two components, three types of noise), SAMD is better than LR and RDBR in four occasions, with the two remaining not showing a statistical significant difference. In all but one case, SAMD is better than MMD (always taking less than a tenth of the time), with the remaining one not showing significant statistical difference.

The obtained waveforms for the noiseless signal, via the four methods, and three more examples on simulated signals can be found in the Supplemental Material.

\subsection{Impedance Pneumography}

\begin{figure}[t]
\begin{center}
\includegraphics[width=\columnwidth]{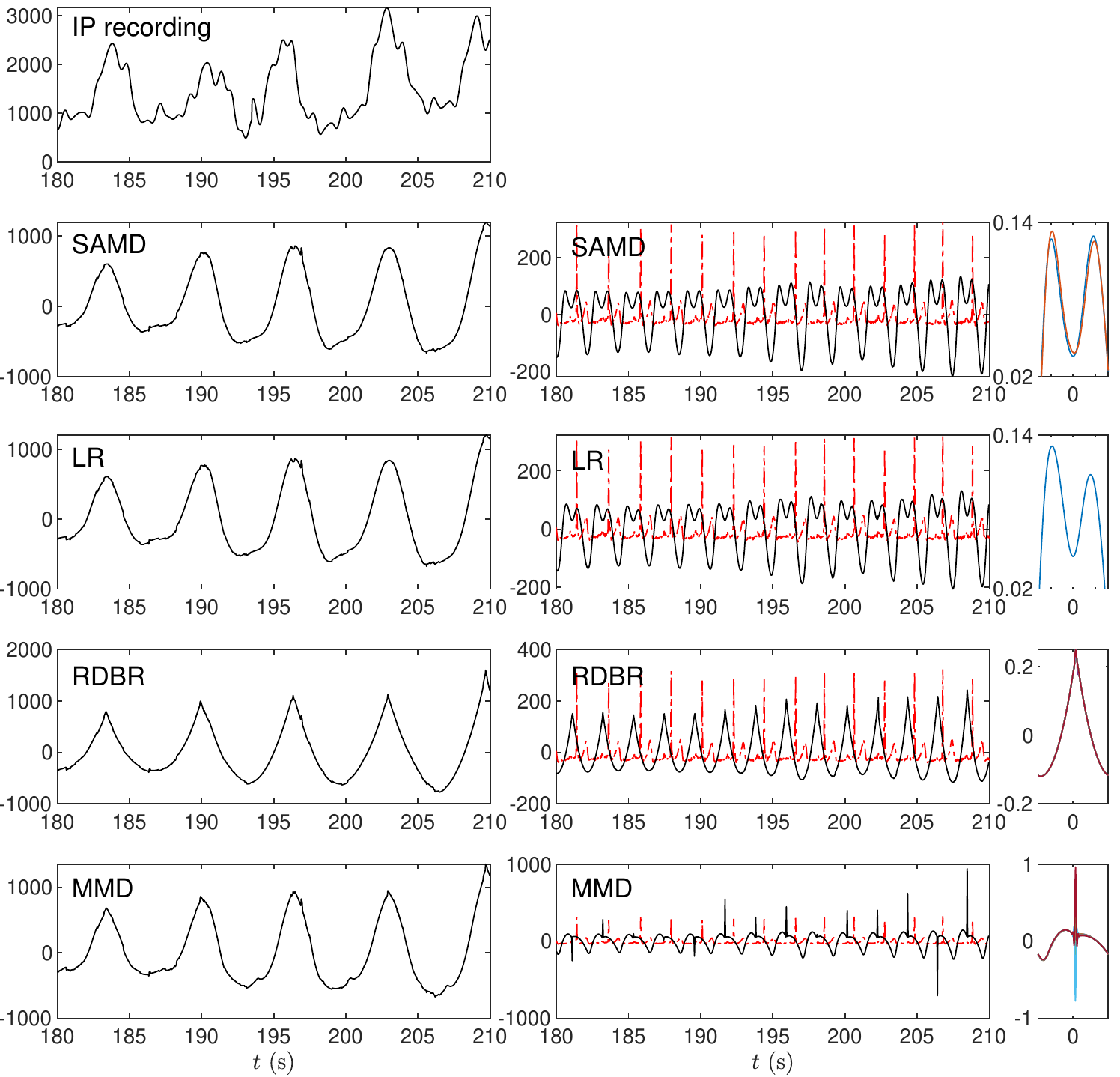}
\end{center}
\cprotect\caption{\textbf{Impedance Pneumography.} First row: 30 seconds of IP recording. Second row: SAMD extracted components and waveforms. Third row: LR extracted components and waveforms. Fourth row: RDBR extracted components and waveforms. Fifth row: MMD extracted components and waveforms. (red dashed-dotted line: ECG signal).}
\label{fig:IP}
\end{figure}

Our first real example is an impedance pneumography (IP) recording \cite{Folke2003}. An IP signal is usually composed of one respiratory component and one cardiac component, called the cardiogenic artifact. Physiologically, the heart rate is faster than the breathing rate, so the condition C4 is satisfied. The IP signal was recorded from patients receiving flexible bronchoscopy examination using the Philips Patient Monitor MP60 at the Chang Gung Memorial Hospital, Linkou, New Taipei, Taiwan. The study protocol was approved by the Chang Gung Medical Foundation Institutional Review Board (No.104-0872C). We applied the four methods to decompose an IP signal of 60 s long. We used a Gaussian window with $\sigma = 0.05$, a maximum jump of 2 Hz and $\Delta = 2$Hz, for the estimation of amplitudes and phases.  $D_1 = 2$ and $D_2 = 5$ were obtained for LR and SAMD, and we used $K = 2$ for SAMD. The results are shown in Fig. \ref{fig:IP}, where only 30 s is shown for the sake of visibility. We show the estimated components, along with the estimated WSF for the cardiac component. For the comparison purpose, we superimposed ECG on top of the extracted cardiac component.

All four methods seem to be able to eliminate the slowly increasing trend present on the signal. SAMD and LR offer similar results, with smooth respiratory components, and cardiac components that match ECG. However, the flexibility of SAMD allows it to capture the WSF change from one cycle to the other. Note that the estimated WSF of the cardiac component by LR is fixed, and has the first bump higher than the second one, while the WSF estimated by SAMD changes along with the signal. We show the first (in blue) and last (in red) cycle (zooming appropriately) to illustrate these changes. The waveforms for both respiratory and cardiac components extracted by RDBR are less physiological. Specifically, the oscillatory patterns of both components are too spiky. As for MMD, the respiratory component is reasonable and comparable with the one extracted by SAMD. However, the cardiac component presents ``spiky'' artifacts, which does not seem to be present in the original data. For the 60-second segment, the computational times were 3.62s, 0.007s, 179.1s, and 34.14s for SAMD, LR, RDBR and MMD respectively.

\subsection{Epileptic newborn electroencephalography}
\begin{figure}[t]
\begin{center}
\includegraphics[width=\columnwidth]{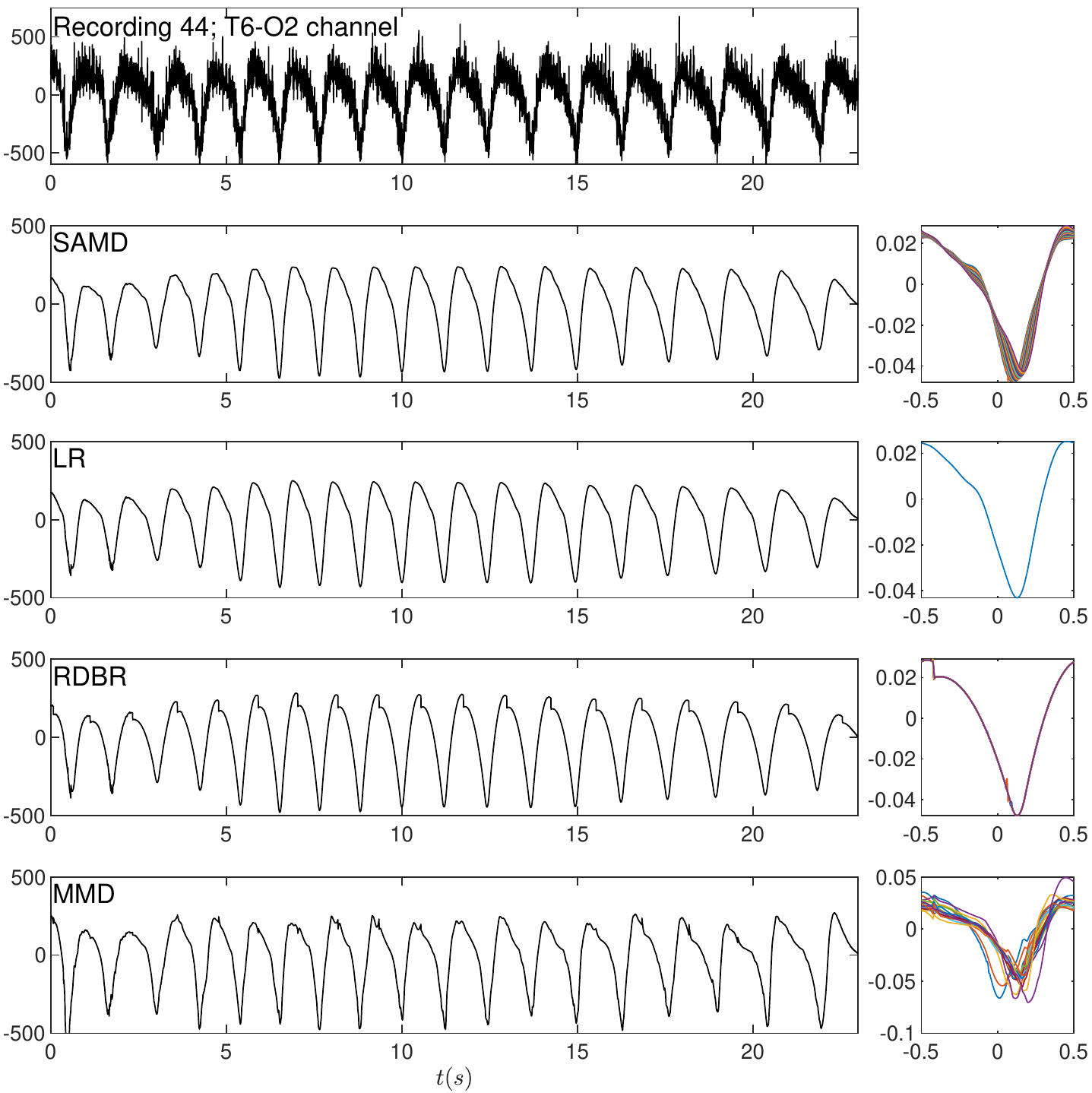}
\end{center}
\cprotect\caption{\textbf{Epileptic newborn electroencephalography.} First row: 23 seconds of channel T6-O2 from recording 44. Second row: SAMD estimated component and waveforms. Third row: LR estimated component and waveform. Fourth row: RDBR estimated component and waveform. Fifth row: MMD estimated component and waveforms.}
\label{fig:EEG}
\end{figure}

We analyzed an EEG recording during the cessation of a widespread seizure discharge with strong muscle artefact, which belongs to a public dataset \cite{stevenson2019dataset}.
We assumed the signal has only one component (i.e. $I = 1$) and applied four methods for the mission of denoising and estimating the WSF. We used a Gaussian window with $\sigma = 0.1$, a maximum jump of 0.5 Hz and $\Delta = 1$Hz, for the estimation of amplitudes and phases. $D_1 = 6$ was obtained for LR and SAMD and we used $K = 1$ for SAMD. The advantages of SAMD are evident. While its waveforms might seem similar to those of LR, SAMD captures the WSF dynamics from one cycle to the other, as can be appreciated on the right column of Fig. \ref{fig:EEG}. On the other hand, RDBR estimates a waveform with a non-smooth behavior, and MMD presents several spiky artifacts, which probably are not physiological. A further exploration from the electrophysiological perspective is needed to further evaluate the performance of these algorithms. The computational times were 1.774s, 0.005s, 5.125s, and 67.788s for SAMD, LR, RDBR and MMD respectively.
An example with an electrocardiogram signal denoising can be found in the Supplemental Material.

\section{Conclusions}\label{sec:Conclusions}
We proposed a novel nonlinear regression algorithm, SAMD, to decompose signals  with time-varying WSFs satisfying the model \eqref{eq:nonlinear_problem}. The flexibility added by allowing the harmonics to be other than integer multiples of a fundamental frequency permits the model to better fit the observed data. The advantages over existing methods, such as RDBR and MMD, are evident both in recovery performance, in the sense of RMSE, and computational times. There are several future research directions we shall consider when applying SAMD to real-world data. First, since long and highly sampled data is getting more popular, a solution to analyze such data like parallelization is needed. A simple solution is the divide and conquer approach -- segment the signal into pieces, run the analysis in parallel, and then concatenate the results. We may need a more sophisticated solution to gain more by parallelization. Real-time implementation is yet another interesting challenge in practice. The main challenge of a real-time implementation of the considered algorithms in this paper is how to handle the boundary effect when a TF analysis tool is applied. To our best knowledge, the most efficient solution so far is based on the forecasting idea \cite{meynard2021efficient}, and it would be interesting to incorporate this idea to SAMD. Another common challenge is dealing with signals with components with crossover IFs; that is, when the condition C4 is not fulfilled. There have been some recent efforts dealing with this challenge for the IF estimation \cite{bruni2018time,bruni2019instantaneous,zhu2020frequency} and decomposition purpose \cite{li2020separation} when IFs crossover, and the Blaschke decomposition can be helpful when IFs do not crossover but are close \cite{coifman2017carrier}. However, these works focus on the sinusoidally oscillatory signal. Another challenge is estimating the number of components, $I$, and $K$ in the model when this information is missing. While the considered peeling scheme for estimating $I$ works well, it can be improved. In the future work we could combine tools like the short-term entropy measure \cite{sucic2011estimating,lerga2017tfd}, but how to accommodate the nontrivial WSF is yet another challenge. Such an estimate is applicable to algorithms like MMD or RDBR, where they assume the knowledge of $I$. The parameter $K$ also needs further exploration. In the current work, $K$ is chosen manually and we found that usually a small $K$, like around $3$, works well. How to find a better solution for these parameters, and the above described challenges, will be the purpose of our future work.

%\bibliographystyle{IEEEtran}
%\bibliography{referencias}

\begin{thebibliography}{10}
\providecommand{\url}[1]{#1}
\csname url@samestyle\endcsname
\providecommand{\newblock}{\relax}
\providecommand{\bibinfo}[2]{#2}
\providecommand{\BIBentrySTDinterwordspacing}{\spaceskip=0pt\relax}
\providecommand{\BIBentryALTinterwordstretchfactor}{4}
\providecommand{\BIBentryALTinterwordspacing}{\spaceskip=\fontdimen2\font plus
\BIBentryALTinterwordstretchfactor\fontdimen3\font minus
  \fontdimen4\font\relax}
\providecommand{\BIBforeignlanguage}[2]{{%
\expandafter\ifx\csname l@#1\endcsname\relax
\typeout{** WARNING: IEEEtran.bst: No hyphenation pattern has been}%
\typeout{** loaded for the language `#1'. Using the pattern for}%
\typeout{** the default language instead.}%
\else
\language=\csname l@#1\endcsname
\fi
#2}}
\providecommand{\BIBdecl}{\relax}
\BIBdecl

\bibitem{akay1998time}
M.~Akay, \emph{Time Frequency and Wavelets in Biomedical Signal
  Processing}.\hskip 1em plus 0.5em minus 0.4em\relax IEEE press series in
  Biomedical Engineering, 1998.

\bibitem{Colominas2014}
M.~A. Colominas, G.~Schlotthauer, and M.~E. Torres, ``Improved complete
  ensemble {EMD}: A suitable tool for biomedical signal processing,''
  \emph{Biomed Signal Proces}, vol.~14, pp. 19--29, 2014.

\bibitem{wu2020current}
H.-T. Wu, ``Current state of nonlinear-type time-frequency analysis and
  applications to high-frequency biomedical signals,'' \emph{Current Opinion in
  Systems Biology}, 2020.

\bibitem{chassande2016ondelettes}
E.~Chassande-Mottin, S.~Jaffard, and Y.~Meyer, ``Des ondelettes pour
  d{\'e}tecter les ondes gravitationnelles,'' 2016.

\bibitem{Pham2017high}
D.-H. Pham and S.~Meignen, ``High-order synchrosqueezing transform for
  multicomponent signals analysis - with an application to gravitational-wave
  signal,'' \emph{IEEE T Signal Proces}, vol.~65, no.~12, pp. 3168--3178, June
  2017.

\bibitem{Daubechies2011synchro}
I.~Daubechies, J.~Lu, and H.-T. Wu, ``Synchrosqueezed wavelet transforms: an
  empirical mode decomposition-like tool,'' \emph{Appl Comput Harmon A},
  vol.~30, no.~2, pp. 243--261, 2011.

\bibitem{Chen_Cheng_Wu:2014}
Y.-C. Chen, M.-Y. Cheng, and H.-T. Wu, ``Nonparametric and adaptive modeling of
  dynamic seasonality and trend with heteroscedastic and dependent errors,''
  \emph{J Roy Stat Soc B}, vol.~76, no.~3, pp. 651--682, 2014.

\bibitem{Gabor:1946}
D.~Gabor, ``Theory of communication. part 1: The analysis of information,''
  \emph{J. Inst. Elec. Engrs. Part III}, vol.~93, pp. 429--441, May 1946.

\bibitem{abramovich2007time}
Y.~I. Abramovich, N.~K. Spencer, and M.~D. Turley, ``Time-varying
  autoregressive (tvar) models for multiple radar observations,'' \emph{IEEE T
  Signal Proces}, vol.~55, no.~4, pp. 1298--1311, 2007.

\bibitem{de2011forecasting}
A.~M. De~Livera, R.~J. Hyndman, and R.~D. Snyder, ``Forecasting time series
  with complex seasonal patterns using exponential smoothing,'' \emph{J Am Stat
  Assoc}, vol. 106, no. 496, pp. 1513--1527, 2011.

\bibitem{lin2019wave}
Y.-T. Lin, J.~Malik, and H.-T. Wu, ``Wave-shape oscillatory model for
  nonstationary periodic time series analysis,'' \emph{Foundations of Data
  Science}, 2021.

\bibitem{Nahon:2000Thesis}
M.~Nahon, ``{Phase Evaluation and Segmentation},'' Ph.D. dissertation, Yale
  University, New Haven, 2000.

\bibitem{HTWu2013}
H.-T. Wu, ``Instantaneous frequency and wave shape functions (i),'' \emph{Appl
  Comput Harmon A}, vol.~35, no.~2, pp. 181--199, 2013.

\bibitem{CYLin2018}
C.-Y. Lin, L.~Su, and H.-T. Wu, ``Wave-shape function analysis,'' \emph{J
  Fourier Anal Appl}, vol.~24, no.~2, pp. 451--505, 2018.

\bibitem{strauss2000relative}
G.~Strauss-Blasche, M.~Moser, M.~Voica, D.~McLeod, N.~Klammer, and W.~Marktl,
  ``Relative timing of inspiration and expiration affects respiratory sinus
  arrhythmia,'' \emph{Clin Exp Pharmacol P}, vol.~27, no.~8, pp. 601--606,
  2000.

\bibitem{tavallali2014extraction}
P.~Tavallali, T.~Y. Hou, and Z.~Shi, ``Extraction of intrawave signals using
  the sparse time-frequency representation method,'' \emph{Multiscale Model
  Sim}, vol.~12, no.~4, pp. 1458--1493, 2014.

\bibitem{Pahlevan_Tavallali_Rinderknecht_Petrasek_Matthews_Hou_Gharib:2014}
N.~M. Pahlevan, P.~Tavallali, D.~G. Rinderknecht, D.~Petrasek, R.~V. Matthews,
  T.~Y. Hou, and M.~Gharib, ``{Intrinsic frequency for a systems approach to
  haemodynamic waveform analysis with clinical applications.}'' \emph{J Roy Soc
  Interface}, vol.~11, no.~98, p. 20140617, 2014.

\bibitem{hou2016extracting}
T.~Y. Hou and Z.~Shi, ``Extracting a shape function for a signal with
  intra-wave frequency modulation,'' \emph{Philos T Roy Soc A}, vol. 374, no.
  2065, p. 20150194, 2016.

\bibitem{Yang2019multiresolution}
H.~Yang, ``Multiresolution mode decomposition for adaptive time series
  analysis,'' \emph{Appl Comput Harmonic A}, 2019.

\bibitem{Xu2018recursive}
J.~Xu, H.~Yang, and I.~Daubechies, ``Recursive diffeomorphism-based regression
  for shape functions,'' \emph{SIAM J Math Anal}, vol.~50, no.~1, pp. 5--32,
  2018.

\bibitem{huang2020scientific}
Y.-C. Huang, A.~Alian, Y.-L. Lo, K.~Shelley, and H.-T. Wu, ``Scientific impact
  of the standard phase reconstruction method and its clinical applications,''
  \emph{bioRxiv}, 2020.

\bibitem{Wu2016modeling}
H.-T. Wu, H.-K. Wu, C.-L. Wang, Y.-L. Yang, W.-H. Wu, T.-H. Tsai, and H.-H.
  Chang, ``Modeling the pulse signal by wave-shape function and analyzing by
  synchrosqueezing transform,'' \emph{PloS one}, vol.~11, no.~6, 2016.

\bibitem{lin2016sleep}
Y.-Y. Lin, H.-T. Wu, C.-A. Hsu, P.-C. Huang, Y.-H. Huang, and Y.-L. Lo, ``Sleep
  apnea detection based on thoracic and abdominal movement signals of wearable
  piezoelectric bands,'' \emph{IEEE J Biomed Health}, vol.~21, no.~6, pp.
  1533--1545, 2016.

\bibitem{Nuttall1966}
A.~Nuttall, ``On the quadrature approximation to the {H}ilbert transform of
  modulated signals,'' \emph{Proceedings of the IEEE}, vol.~54, pp. 1458--1459,
  1966.

\bibitem{Cohen1995}
L.~Cohen, \emph{Time-frequency analysis}.\hskip 1em plus 0.5em minus
  0.4em\relax Prentice Hall, 1995, vol.~1, no. 995,299.

\bibitem{Carmona1999}
R.~Carmona, W.~L. Hwang, and B.~Torr{\'e}sani, ``Multiridge detection and
  time-frequency reconstruction,'' \emph{IEEE T Signal Proces}, vol.~47, no.~2,
  pp. 480--492, 1999.

\bibitem{Wu2011adaptive}
H.-T. Wu, ``Adaptive analysis of complex data sets,'' Ph.D. dissertation,
  Princeton University, 2011.

\bibitem{Oberlin2014}
T.~Oberlin, S.~Meignen, and V.~Perrier, ``The {F}ourier-based synchrosqueezing
  transform,'' in \emph{2014 IEEE Int Conf Acoust Spee (ICASSP)}.\hskip 1em
  plus 0.5em minus 0.4em\relax IEEE, 2014, pp. 315--319.

\bibitem{Oberlin2015}
------, ``Second-order synchrosqueezing transform or invertible reassignment?
  {T}owards ideal time-frequency representations,'' \emph{IEEE T Signal
  Proces}, vol.~63, no.~5, pp. 1335--1344, 2015.

\bibitem{fourer2017chirp}
D.~Fourer, F.~Auger, K.~Czarnecki, S.~Meignen, and P.~Flandrin, ``Chirp rate
  and instantaneous frequency estimation: application to recursive vertical
  synchrosqueezing,'' \emph{IEEE Signal Proc Let}, vol.~24, no.~11, pp.
  1724--1728, 2017.

\bibitem{Auger1995}
F.~Auger and P.~Flandrin, ``Improving the readability of time-frequency and
  time-scale representations by the reassignment method,'' \emph{IEEE T Signal
  Proces}, vol.~43, no.~5, pp. 1068--1089, 1995.

\bibitem{Flandrin2018explorations}
P.~Flandrin, \emph{Explorations in time-frequency analysis}.\hskip 1em plus
  0.5em minus 0.4em\relax Cambridge University Press, 2018.

\bibitem{Meignen2017}
S.~Meignen, D.-H. Pham, and S.~McLaughlin, ``On demodulation, ridge detection,
  and synchrosqueezing for multicomponent signals,'' \emph{IEEE T Signal
  Proces}, vol.~65, no.~8, pp. 2093--2103, 2017.

\bibitem{Colominas2020}
M.~A. Colominas, S.~Meignen, and D.~H. Pham, ``Fully adaptive ridge detection
  based on {STFT} phase information,'' \emph{IEEE Signal Proc Let}, vol.~27,
  pp. 620--624, 2020.

\bibitem{Auger2013}
F.~Auger, P.~Flandrin, Y.-T. Lin, S.~McLaughlin, S.~Meignen, T.~Oberlin, and
  H.-T. Wu, ``Time-frequency reassignment and synchrosqueezing: An overview,''
  \emph{IEEE Signal Proc Mag}, vol.~30, no.~6, pp. 32--41, 2013.

\bibitem{sourisseau2019inference}
M.~Sourisseau, H.-T. Wu, and Z.~Zhou, ``Inference of synchrosqueezing
  transform--toward a unified statistical analysis of nonlinear-type
  time-frequency analysis,'' \emph{arXiv preprint arXiv:1904.09534}, 2019.

\bibitem{ikram2001estimation}
M.~Z. Ikram and G.~T. Zhou, ``Estimation of multicomponent polynomial phase
  signals of mixed orders,'' \emph{Signal Process}, vol.~81, no.~11, pp.
  2293--2308, 2001.

\bibitem{Huang1998}
N.~E. Huang, Z.~Shen, S.~R. Long, M.~C. Wu, H.~H. Shih, Q.~Zheng, N.-C. Yen,
  C.~C. Tung, and H.~H. Liu, ``The empirical mode decomposition and the
  {H}ilbert spectrum for nonlinear and non-stationary time series analysis,''
  in \emph{P Roy Soc Lond A Mat}, vol. 454, no. 1971, 1998, pp. 903--995.

\bibitem{lerga2011improved}
J.~Lerga, V.~Sucic, and B.~Boashash, ``An improved method for nonstationary
  signals components extraction based on the {ICI} rule,'' in \emph{Int
  Workshop on Systems, Signal Processing and their Applications, WOSSPA}.\hskip
  1em plus 0.5em minus 0.4em\relax IEEE, 2011, pp. 307--310.

\bibitem{khan2012instantaneous}
N.~A. Khan and B.~Boashash, ``Instantaneous frequency estimation of
  multicomponent nonstationary signals using multiview time-frequency
  distributions based on the adaptive fractional spectrogram,'' \emph{IEEE
  Signal Proc Let}, vol.~20, no.~2, pp. 157--160, 2012.

\bibitem{lerga2011efficient}
J.~Lerga, V.~Sucic, and B.~Boashash, ``An efficient algorithm for instantaneous
  frequency estimation of nonstationary multicomponent signals in low {SNR},''
  \emph{EURASIP J Adv Sig Pr}, vol. 2011, pp. 1--16, 2011.

\bibitem{Yildirim2002}
E.~A. Yildirim and S.~J. Wright, ``Warm-start strategies in interior-point
  methods for linear programming,'' \emph{SIAM J Optimiz}, vol.~12, no.~3, pp.
  782--810, 2002.

\bibitem{dumouchel1989integrating}
W.~Dumouchel, F.~O'Brien \emph{et~al.}, ``Integrating a robust option into a
  multiple regression computing environment,'' in \emph{Computer science and
  statistics: Proceedings of the 21st symposium on the interface}.\hskip 1em
  plus 0.5em minus 0.4em\relax American Statistical Association Alexandria, VA,
  1989, pp. 297--302.

\bibitem{holland1977robust}
P.~W. Holland and R.~E. Welsch, ``Robust regression using iteratively
  reweighted least-squares,'' \emph{Commun Stat A-Theor}, vol.~6, no.~9, pp.
  813--827, 1977.

\bibitem{Baraniuk2001}
R.~G. Baraniuk, P.~Flandrin, A.~J. Janssen, and O.~J. Michel, ``Measuring
  time-frequency information content using the {R}{\'e}nyi entropies,''
  \emph{IEEE T Inform Theory}, vol.~47, no.~4, pp. 1391--1409, 2001.

\bibitem{Meignen2020}
S.~Meignen, M.~A. Colominas, and D.-H. Pham, ``On the use of {R}{\'e}nyi
  entropy for optimal window size computation in the short-time {F}ourier
  transform,'' in \emph{2014 IEEE Int Conf Acoust Spee (ICASSP)}.\hskip 1em
  plus 0.5em minus 0.4em\relax IEEE, 2020, pp. 5830--5834.

\bibitem{Kavalieris1994}
L.~Kavalieris and E.~Hannan, ``Determining the number of terms in a
  trigonometric regression,'' \emph{J Time Ser Anal}, vol.~15, no.~6, pp.
  613--625, 1994.

\bibitem{Ruiz2020}
J.~Ruiz and M.~A. Colominas, ``Wave-shape function model order estimation by
  trigonometric regression,'' \emph{submitted}, 2020.

\bibitem{Folke2003}
M.~Folke, L.~Cernerud, M.~Ekstr{\"o}m, and B.~H{\"o}k, ``Critical review of
  non-invasive respiratory monitoring in medical care,'' \emph{Med Biol Eng
  Comput}, vol.~41, no.~4, pp. 377--383, 2003.

\bibitem{stevenson2019dataset}
N.~Stevenson, K.~Tapani, L.~Lauronen, and S.~Vanhatalo, ``A dataset of neonatal
  {EEG} recordings with seizure annotations,'' \emph{Scientific Data}, vol.~6,
  no.~1, pp. 1--8, 2019.

\bibitem{meynard2021efficient}
A.~Meynard and H.-T. Wu, ``An efficient forecasting approach to reduce boundary
  effects in real-time time-frequency analysis,'' \emph{IEEE T Signal Proces},
  vol.~69, pp. 1653--1663, 2021.

\bibitem{bruni2018time}
V.~Bruni, M.~Tartaglione, and D.~Vitulano, ``On the time-frequency reassignment
  of interfering modes in multicomponent fm signals,'' in \emph{26th Eur Signal
  Pr Conf (EUSIPCO)}.\hskip 1em plus 0.5em minus 0.4em\relax IEEE, 2018, pp.
  722--726.

\bibitem{bruni2019instantaneous}
------, ``Instantaneous frequency modes separation via a spectrogram-radon
  based approach,'' in \emph{11th Int Symposium on Image and Signal Processing
  and Analysis (ISPA)}.\hskip 1em plus 0.5em minus 0.4em\relax IEEE, 2019, pp.
  347--351.

\bibitem{zhu2020frequency}
X.~Zhu, H.~Yang, Z.~Zhang, J.~Gao, and N.~Liu, ``Frequency-chirprate
  reassignment,'' \emph{Digit Signal Process}, vol. 104, p. 102783, 2020.

\bibitem{li2020separation}
L.~Li, N.~Han, Q.~Jiang, and C.~K. Chui, ``A separation method for
  multicomponent nonstationary signals with crossover instantaneous
  frequencies,'' \emph{arXiv preprint arXiv:2010.01498}, 2020.

\bibitem{coifman2017carrier}
R.~R. Coifman, S.~Steinerberger, and H.-t. Wu, ``Carrier frequencies,
  holomorphy, and unwinding,'' \emph{SIAM J Math Anal}, vol.~49, no.~6, pp.
  4838--4864, 2017.

\bibitem{sucic2011estimating}
V.~Sucic, N.~Saulig, and B.~Boashash, ``Estimating the number of components of
  a multicomponent nonstationary signal using the short-term time-frequency
  {R}{\'e}nyi entropy,'' \emph{EURASIP J Adv Sig Pr}, vol. 2011, no.~1, pp.
  1--11, 2011.

\bibitem{lerga2017tfd}
J.~Lerga, N.~Saulig, R.~Lerga, and I.~{\v{S}}tajduhar, ``{TFD} thresholding in
  estimating the number of {EEG} components and the dominant {IF} using the
  short-term {R}{\'e}nyi entropy,'' in \emph{Proc 10th Int Symposium on Image
  and Signal Processing and Analysis}.\hskip 1em plus 0.5em minus 0.4em\relax
  IEEE, 2017, pp. 80--85.

\end{thebibliography}

\begin{thebibliography}{1}
\providecommand{\url}[1]{#1}
\csname url@samestyle\endcsname
\providecommand{\newblock}{\relax}
\providecommand{\bibinfo}[2]{#2}
\providecommand{\BIBentrySTDinterwordspacing}{\spaceskip=0pt\relax}
\providecommand{\BIBentryALTinterwordstretchfactor}{4}
\providecommand{\BIBentryALTinterwordspacing}{\spaceskip=\fontdimen2\font plus
\BIBentryALTinterwordstretchfactor\fontdimen3\font minus
  \fontdimen4\font\relax}
\providecommand{\BIBforeignlanguage}[2]{{%
\expandafter\ifx\csname l@#1\endcsname\relax
\typeout{** WARNING: IEEEtran.bst: No hyphenation pattern has been}%
\typeout{** loaded for the language `#1'. Using the pattern for}%
\typeout{** the default language instead.}%
\else
\language=\csname l@#1\endcsname
\fi
#2}}
\providecommand{\BIBdecl}{\relax}
\BIBdecl

\bibitem{Baraniuk2001}
R.~G. Baraniuk, P.~Flandrin, A.~J. Janssen, and O.~J. Michel, ``Measuring
  time-frequency information content using the {R}{\'e}nyi entropies,''
  \emph{IEEE T Inform Theory}, vol.~47, no.~4, pp. 1391--1409, 2001.

\bibitem{Meignen2020}
S.~Meignen, M.~A. Colominas, and D.-H. Pham, ``On the use of {R}{\'e}nyi
  entropy for optimal window size computation in the short-time {F}ourier
  transform,'' in \emph{2014 IEEE Int Conf Acoust Spee (ICASSP)}.\hskip 1em
  plus 0.5em minus 0.4em\relax IEEE, 2020, pp. 5830--5834.

\bibitem{Ruiz2020}
J.~Ruiz and M.~A. Colominas, ``Wave-shape function model order estimation by
  trigonometric regression,'' \emph{submitted}, 2020.

\bibitem{iyengar1996age}
N.~Iyengar, C.~Peng, R.~Morin, A.~L. Goldberger, and L.~A. Lipsitz,
  ``Age-related alterations in the fractal scaling of cardiac interbeat
  interval dynamics,'' \emph{American Journal of Physiology-Regulatory,
  Integrative and Comparative Physiology}, vol. 271, no.~4, pp. R1078--R1084,
  1996.

\bibitem{Physionet}
A.~L. Goldberger, L.~A. Amaral, L.~Glass, J.~M. Hausdorff, P.~C. Ivanov, R.~G.
  Mark, J.~E. Mietus, G.~B. Moody, C.-K. Peng, and H.~E. Stanley, ``Physiobank,
  physiotoolkit, and physionet: components of a new research resource for
  complex physiologic signals,'' \emph{circulation}, vol. 101, no.~23, pp.
  e215--e220, 2000.

\end{thebibliography}

\section*{Supplemental Material}

\section{The first simulated signal (noiseless case)}

\begin{figure}[h!]
\begin{center}
\includegraphics[width=\columnwidth]{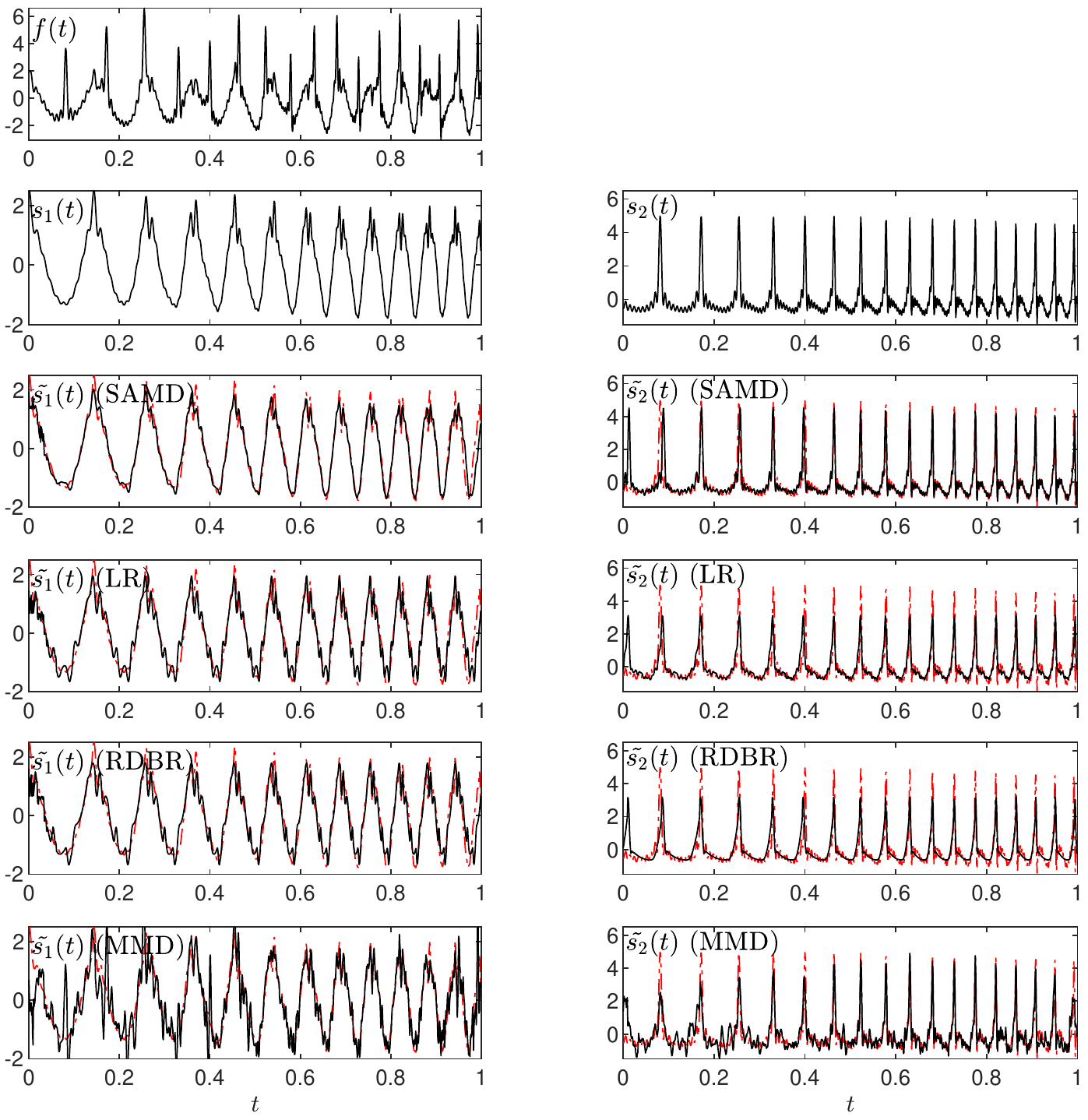}
\end{center}
\caption{\textbf{Simulated signal (noiseless).} First row: analyzed signal $f(t)$ from Eqs. \eqref{eq:s1_6}, and \eqref{eq:s2_6}. Second row: true components. Third row: extracted modes with our proposal SAMD. Fourth row: extracted components with LR. Fifth row: extracted components with RDBR. Sixth row: extracted components with MMD. (black solid line: estimated component; red dashed line: true components).}
\label{fig:06}
\end{figure}

Figure \ref{fig:06} presents the decomposition results of the simulated signal shown in the main manuscript when noise does not exist. We recall the simulation here for the sake of completeness: $f(t) = s_1(t) + s_2(t)$, where
\begin{equation}\label{eq:s1_6}
\begin{aligned}
s&_1(t) = 1.5 \cos(\phi_{11}(t)) + 0.25\cos(2.05 \phi_{11}(t)) \\
&+ \sum_{p = 3}^{10} 0.1 \cos((p + 0.05) \phi_{11}(t) + 0.01\phi_{11}^2(t)),
\end{aligned}
\end{equation}
where $\phi_{11}(t) = 2\pi 6 t + 2\pi 6 t^2$, and
\begin{equation}\label{eq:s2_6}
s_2(t) = \cos(\phi_{21}(t)) \!+\! \sum_{p = 2}^{10} \frac{\cos((p+0.01)\phi_{21}(t))}{\sqrt{p}},
\end{equation}
where $\phi_{21}(t) = 2\pi 10 t + 2\pi 7 t^2 + 0.5 \cos(2 \pi t)$. The signals are defined for $t\in[0,1]$, and sampled at 1000 Hz.

\section{The second simulated signal}

We consider another simulated example with two oscillatory components: $f(t) = s_1(t) + s_2(t)$, where
\begin{equation}\label{eq:s1}
s_1(t) = \sum_p e^{-2\times10^5 (t-t_p)^2},
\end{equation}
 with $t_p$ such that $\phi_{11}(t_p) = 2 \pi p, p\in \mathbb{Z}$ and $\phi_{11}(t) = 2 \pi 20 t + 2 \cos(4 \pi t)$, and
\begin{equation}\label{eq:s2}
s_2(t) = 0.5\cos(\phi_{21}(t)) + 0.375 \cos(2.05 \phi_{21}(t)),
\end{equation}
with $\phi_{21}(t) = 2\pi 10 t + 2 \pi 5 t^2$. For $s_1$, we subtracted its mean to make it a zero-mean signal. The signals are defined for $t\in[0,1]$, and sampled at 1000 Hz.

Figure \ref{fig:01} presents the results for the decomposition of the second simulated signal without noise contamination. We used the parameters $D_1 = 10$, $D_2 = 2$, for both LR and SAMD methods, and $K = 3$ following the same procedure detailed in the main article. The errors can be found in Table I, with suggests that SAMD provides a better result.

\begin{figure}[t]
\begin{center}
\includegraphics[width=\columnwidth]{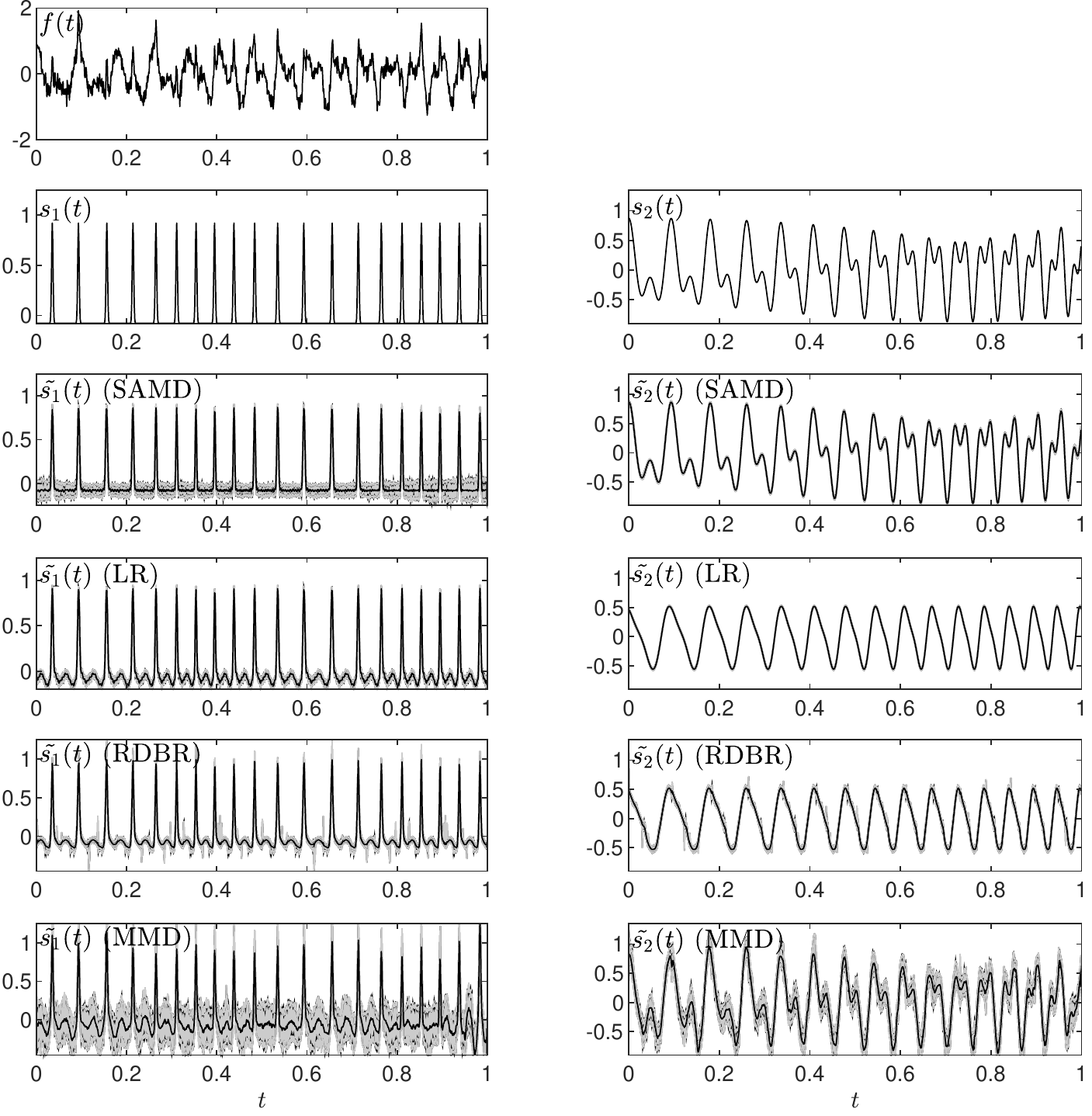}
\end{center}
\caption{\textbf{Second simulated signal (noisy signals at 10 dB; 100 realizations).} First row: a typical noisy example of signal $f(t)$ from Eqs. \eqref{eq:s1} and \eqref{eq:s2}. Second row: true components. Third row: extracted modes with our proposal SAMD. Fourth row: extracted components with LR. Fifth row: extracted components with RDBR. Sixth row: extracted components with MMD. (black solid line: mean estimated components; shaded gray area: 95\% confidence interval).}
\label{fig:01_ruido}
\end{figure}

We tested the robustness to noise by applying different algorithms to 100 realizations of noisy copies of the signal at 10 dB. The noise is Gaussian white. We present the results in Fig. \ref{fig:01_ruido}, where we show the mean and the 95\% confidence interval. The means and standard deviations of the RMSE are presented on Table I, along with the computational times.

\begin{figure}[t]
\begin{center}
\includegraphics[width=\columnwidth]{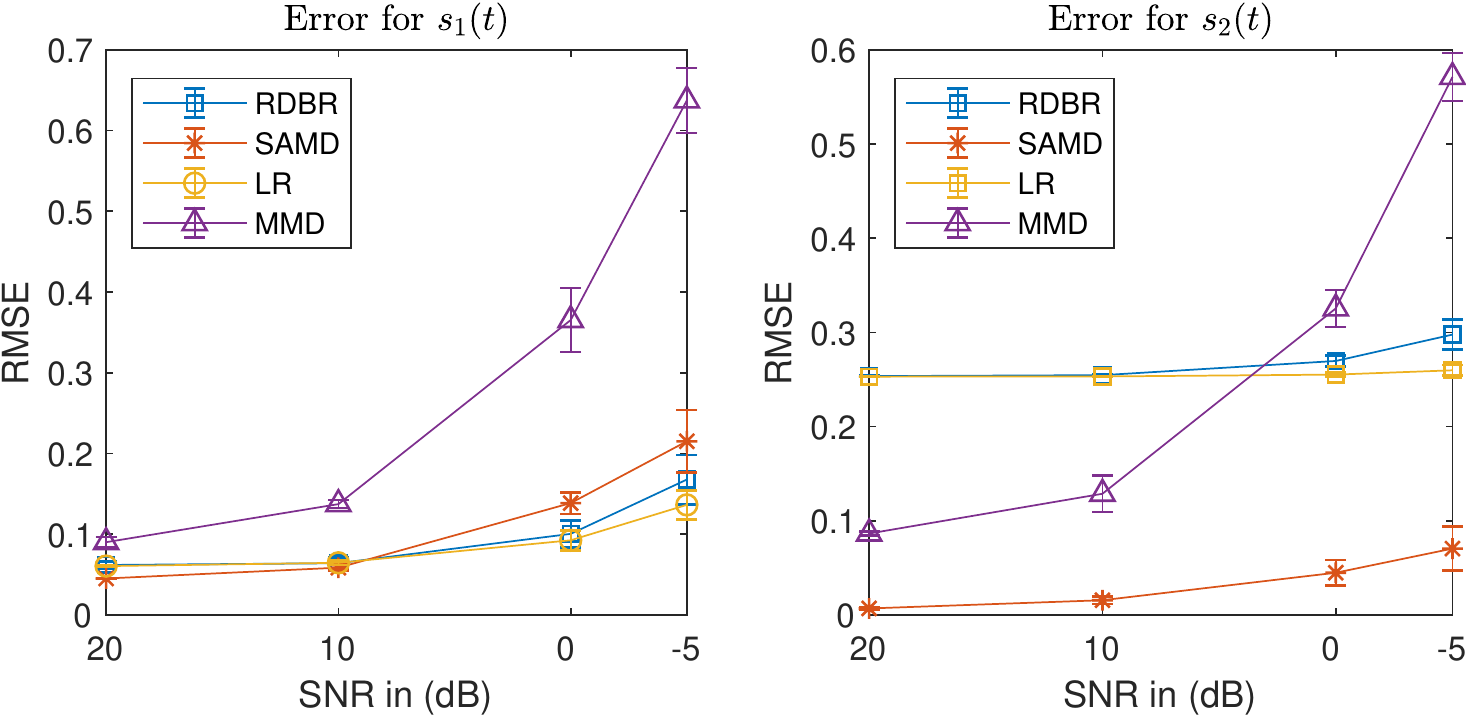}
\end{center}
\caption{\textbf{Errors for the second simulated signal (different SNRs; 50 realizations).} Left: mean errors and standard deviations for $s_1(t)$. Right: mean errors and standard deviations for $s_2(t)$. For MMD method, we considered only those decompositions that converged (between 44 and 48 times out of the 50 realizations, depending on the input SNR).}
\label{fig:01_varias}
\end{figure}

Figure \ref{fig:01_varias} presents mean RMSEs (and its standard deviations) of the decomposition at different SNRs (50 realizations). While SAMD showed the best performance across the different methods, it also presented an acceptable robustness since its behavior does not worsen as that of MMD when the input SNR decreases.

\begin{table}\label{table:01}
\caption{Errors and computation times for second simulated signal from Eqs. \eqref{eq:s1} and \eqref{eq:s2}. *Out of the 100 realizations, MMD converged on 92 occasions.}
\begin{center}
\begin{tabular}{lcccc}
\hline
   & Noiseless & \multicolumn{3}{c}{Noisy (10 dB; 100 realizations)} \\
   & RMSE  & mean(RMSE) & std(RMSE)& mean time (s) \\
\hline
$s_1$ (SAMD) & 0.0482 &  0.0586  & 0.0043 &\multirow{2}{*}{12.7855} \\
$s_2$ (SAMD) & 0.0002 & 0.0152 & 0.0041 &  \\
\hline
$s_1$ (LR) & 0.0602  &  0.0641  & 0.0032 &\multirow{2}{*}{0.0012}\\
$s_2$ (LR) & 0.253 & 0.2532 & 0.0002 &  \\
\hline
$s_1$ (RDBR) & 0.0626  &  0.0638  & 0.0080 &\multirow{2}{*}{16.9325} \\
$s_2$ (RDBR) & 0.2532 & 0.2547 & 0.001 &  \\
\hline
$s_1$ (MMD*) & 0.0820 &  0.1495 & 0.0806 &\multirow{2}{*}{270}  \\
$s_2$ (MMD*) & 0.0821 &  0.1322 & 0.0375 &  \\
\end{tabular}
\end{center}
\end{table}

\begin{figure}[t]
\begin{center}
\includegraphics[width=\columnwidth]{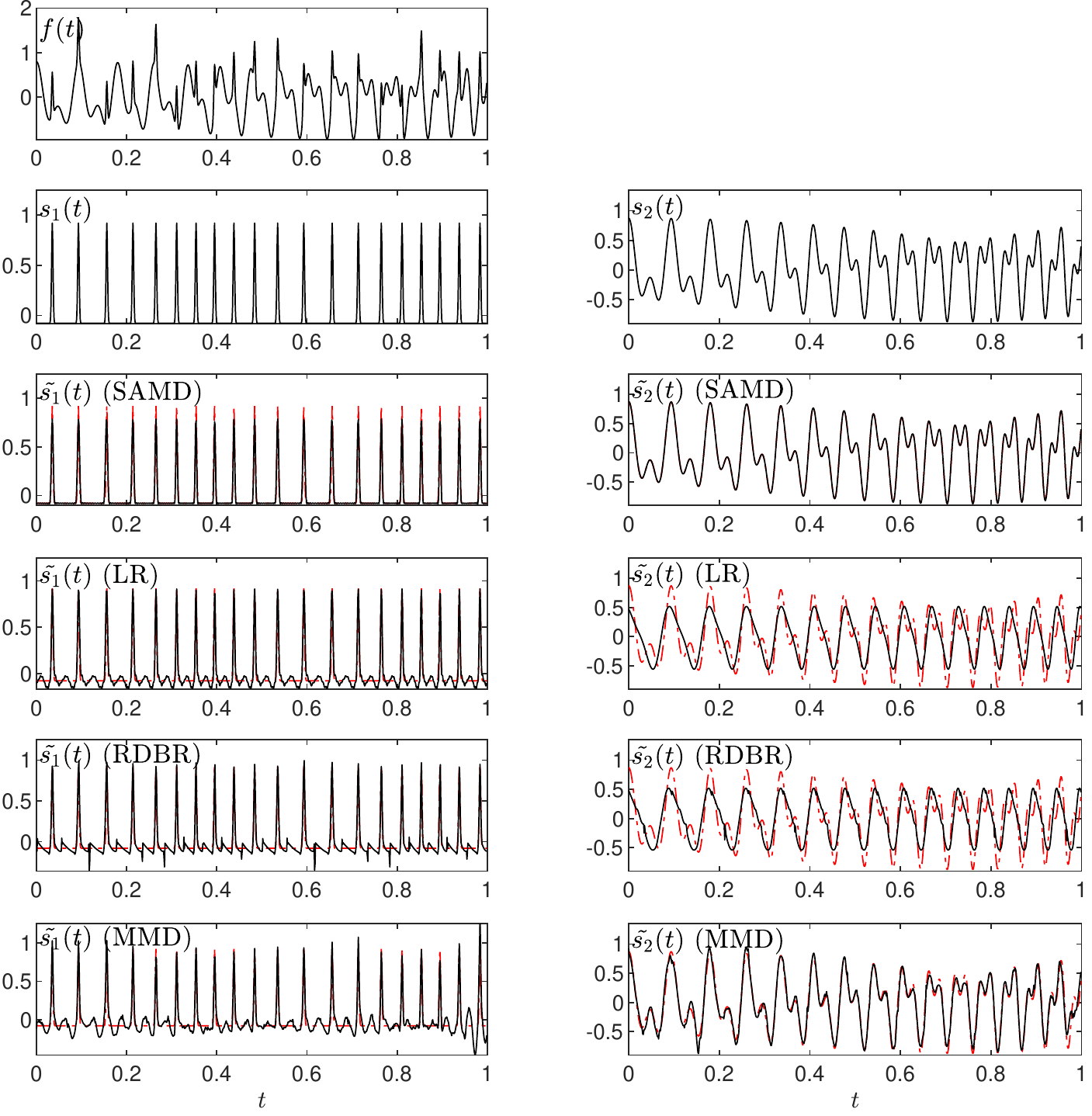}
\end{center}
\caption{\textbf{Second simulated signal (noiseless).} First row: analyzed signal $f(t)$ from Eqs. \eqref{eq:s1} and \eqref{eq:s2}. Second row: true components. Third row: extracted modes with our proposal SAMD. Fourth row: extracted components with LR. Fifth row: extracted components with RDBR. Sixth row: extracted components with MMD. (black solid line: estimated component; red dashed line: true components).}
\label{fig:01}
\end{figure}

\section{The third simulated signal}

As a third example, we decomposed $f(t) = s_1(t) + s_2(t)$, where
\begin{equation}\label{eq:s1_2}
s_1(t) = \cos(\phi_{11}(t)) + 0.5\cos(1.9\phi_{11} + 0.01\phi_{11}^2(t)),
\end{equation}
 with $\phi_{11}(t) = 2\pi 3 t + 2\pi t^2$, and
\begin{equation}\label{eq:s2_2}
s_2(t) = \cos(\phi_{21}(t)) + \sum_{p = 2}^5 \frac{1}{p} \cos((p+0.01)\phi_{21}(t)),
\end{equation}
 with $\phi_{21}(t) = 2\pi 10 t + 2\pi 10 t^2$. As before, the signals are defined for $t \in [0,1]$, and sampled at 1000 Hz.

This time, in order to have a complete picture of the different methods, we incorporated the estimations of $\phi_{11}(t)$ and $\phi_{21}(t)$. As explained in the paper, we used second-order SST, and ridge detection algorithm for all four methods. We used a Gaussian window $g(t) = \sigma e^{-\frac{\pi t^2}{\sigma^2}}$, with $\sigma = 0.45$ (which minimizes the criterion of the R\'enyi entropy \cite{Baraniuk2001,Meignen2020}), for the ridge detection we allowed a maximum jump of 2 Hz, and for estimating the complex components we used $\Delta = 1$Hz.

In order to avoid possible boundary effects due to ridge estimation, we computed the RMSEs on the interval $(0.1,0.9)$.

\begin{figure}[t]
\begin{center}
\includegraphics[width=\columnwidth]{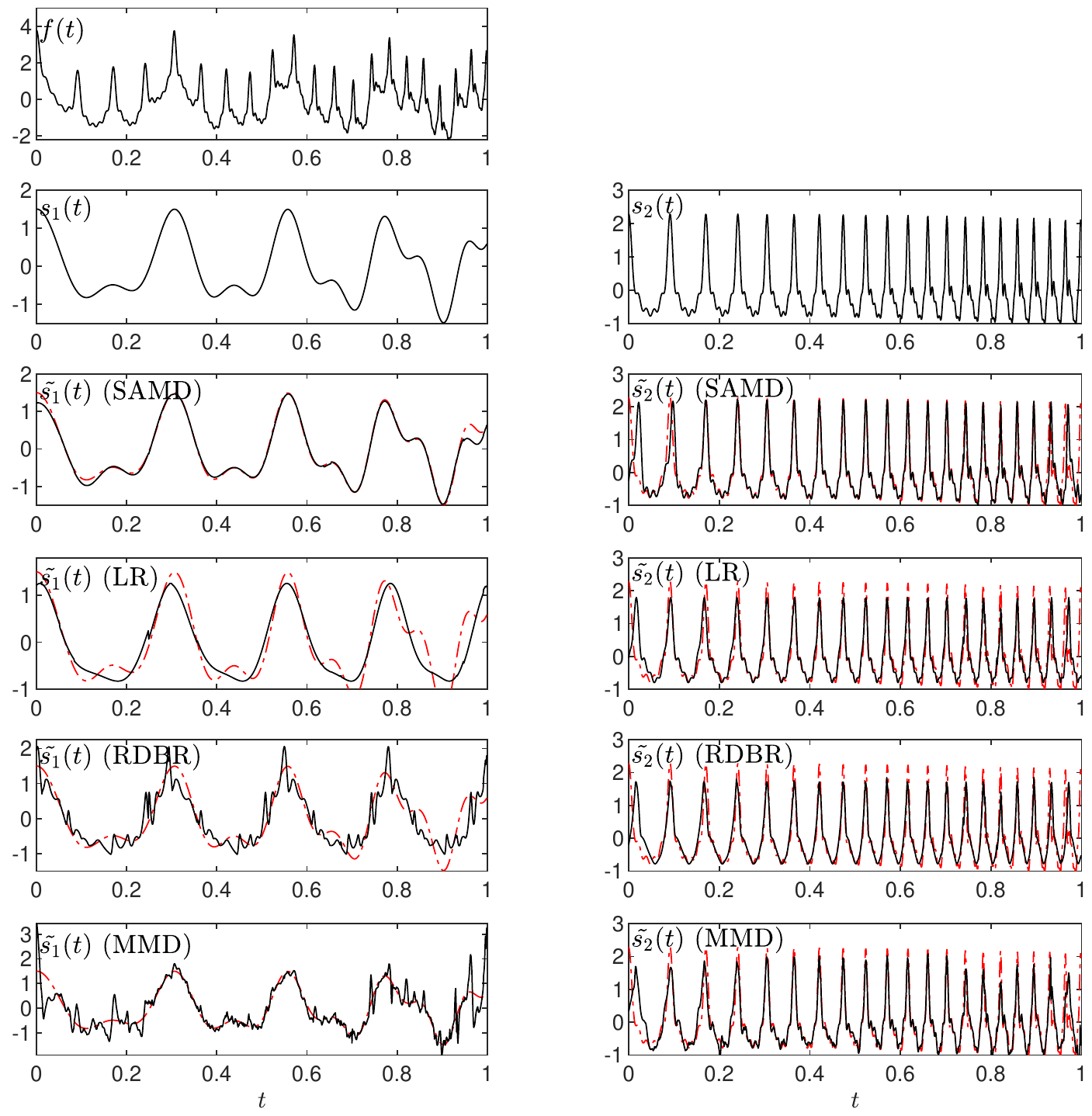}
\end{center}
\caption{\textbf{Third simulated signal (noiseless).} First row: analyzed signal $f(t)$ from Eqs. \eqref{eq:s1_2} and \eqref{eq:s2_2}. Second row: true components. Third row: extracted modes with our proposal SAMD. Fourth row: extracted components with LR. Fifth row: extracted components with RDBR. Sixth row: extracted components with MMD. (black solid line: estimated component; red dashed line: true components).}
\label{fig:02}
\end{figure}

The results for the four methods are presented on Fig. \ref{fig:02}. We used the parameters $D_1 = 2$, $D_2 =
5$, for both LR and SAMD methods, and $K = 3$ for the phases estimations in SAMD, following the same procedure detailed in the main article. The errors (measured as the RMSE on the mentioned interval) can be found in Table II, with clear advantages for our proposal.

\begin{figure}[t]
\begin{center}
\includegraphics[width=\columnwidth]{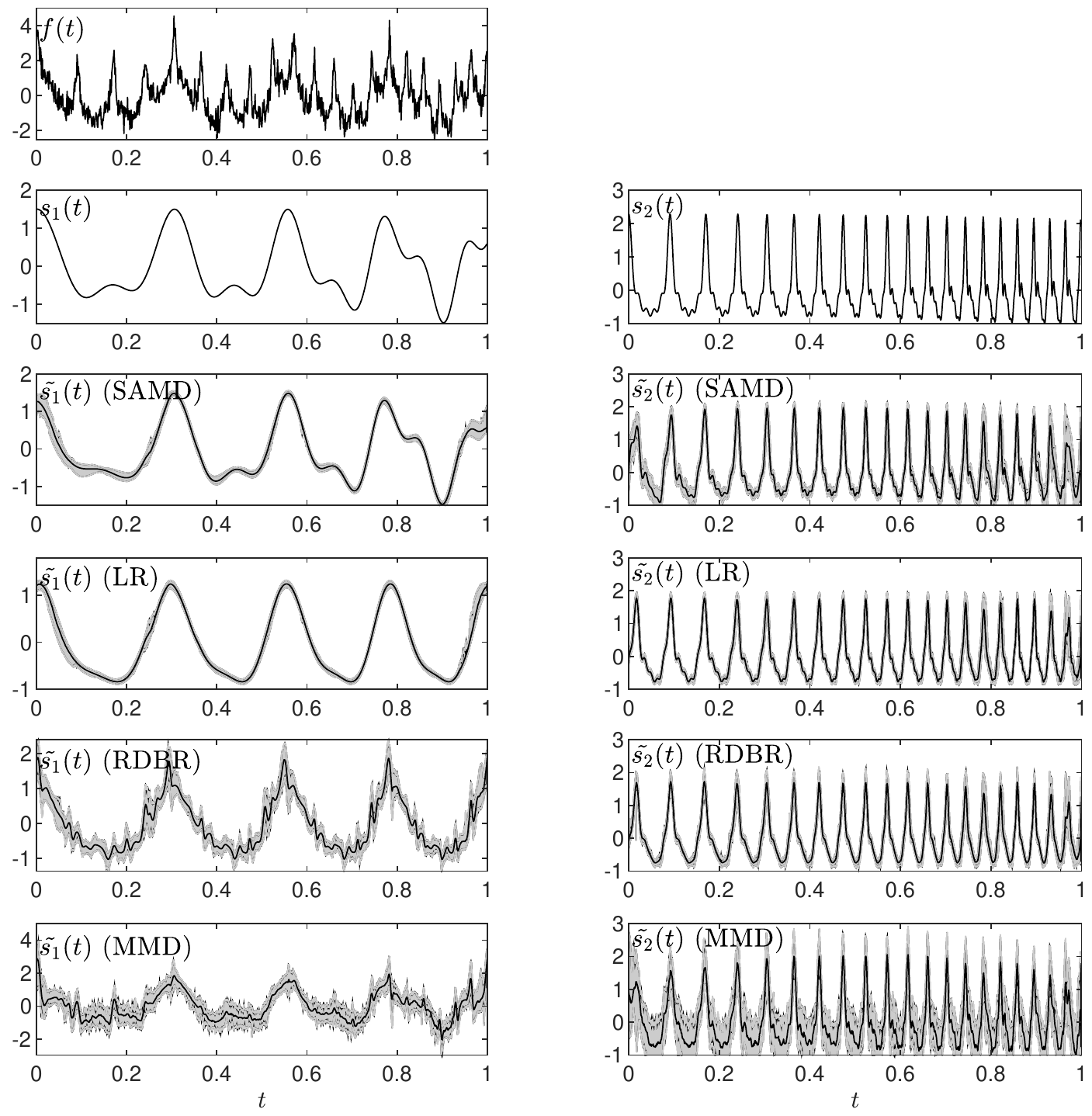}
\end{center}
\caption{\textbf{Third simulated signal (noisy signals at 10 dB; 100 realizations).} First row: a typical noisy example of signal $f(t)$ from Eqs. \eqref{eq:s1_2} and \eqref{eq:s2_2}. Second row: true components. Third row: extracted modes with our proposal SAMD. Fourth row: extracted components with LR. Fifth row: extracted components with RDBR. Sixth row: extracted components with MMD. (black solid line: mean estimated components; shaded gray area: 95\% confidence interval).}
\label{fig:02_ruido}
\end{figure}

Noise robustness was tested by decomposing 100 realizations of noisy copies of the signal at 10 dB. The results can be appreciated on Fig. \ref{fig:02_ruido}, where we show the mean and the 95\% confidence interval. The means and standard deviations of the RMSE are presented on Table \ref{table:02}, along with the computational times.

\begin{figure}[t]
\begin{center}
\includegraphics[width=\columnwidth]{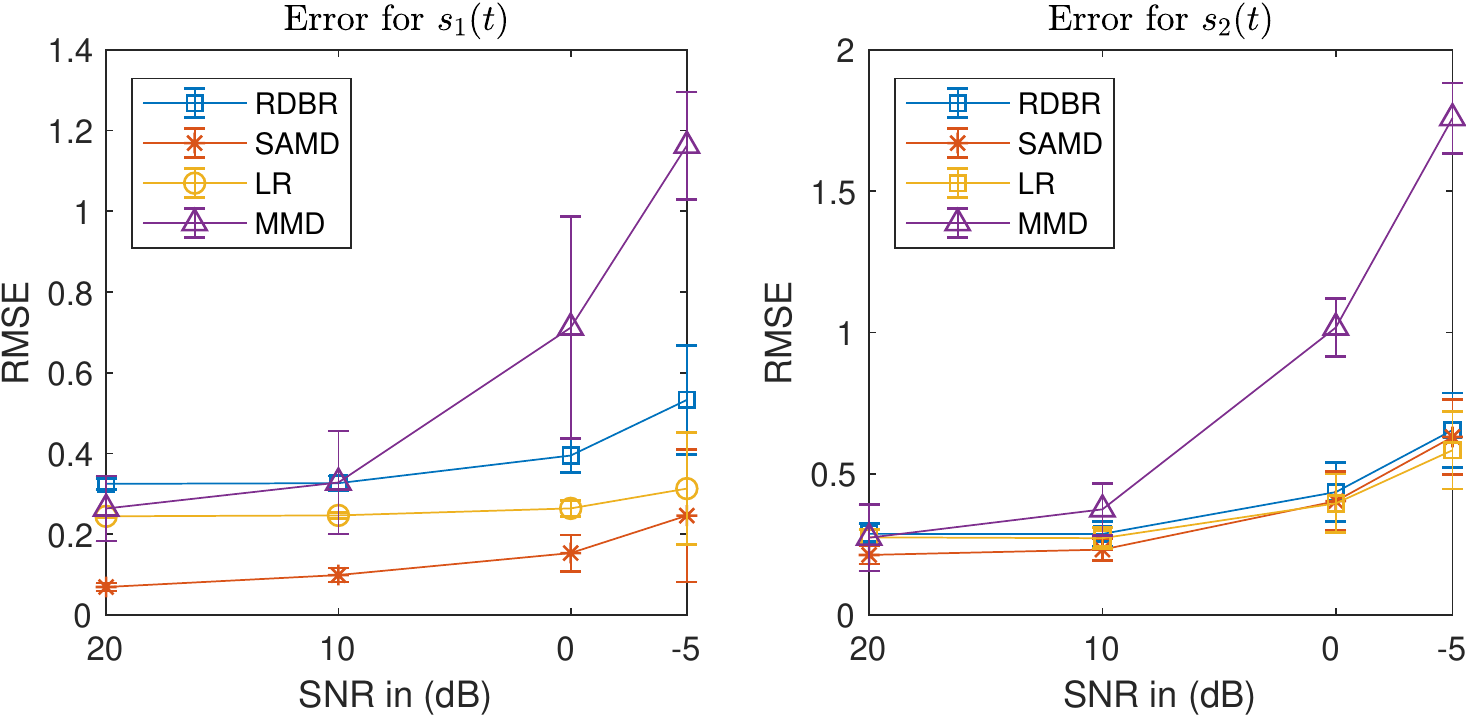}
\end{center}
\caption{\textbf{Errors for the third simulated signal (different SNRs; 50 realizations).} Left: mean errors and standard deviations for $s_1(t)$. Right: mean errors and standard deviations for $s_2(t)$. For MMD method, we considered only those decompositions that converged (between 43 and 49 times out of the 50 realizations, depending on the input SNR).}
\label{fig:02_varias}
\end{figure}

Figure \ref{fig:02_varias} presents mean RMSEs (and its standard deviations) of the results when decomposing the signal at different SNRs (50 realizations). While SAMD showed the best performance among different methods, it also presented an acceptable robustness. Particularly, its behavior does not worsen as that of MMD when the input SNR decreases.

\begin{table}\label{table:02}
\caption{Errors and computation times for third simulated signal from Eqs. \eqref{eq:s1_2} and \eqref{eq:s2_2}. *Out of the 100 realizations, MMD converged on 95 occasions.}
\begin{center}
\begin{tabular}{lcccc}
\hline
   & Noiseless & \multicolumn{3}{c}{Noisy (10 dB; 100 realizations)} \\
             & RMSE   & mean(RMSE) & std(RMSE)& mean time (s) \\
\hline
$s_1$ (SAMD) & 0.0432 &  0.1070  & 0.0147 &\multirow{2}{*}{7.1416} \\
$s_2$ (SAMD) & 0.1645 & 0.2503   & 0.0367 &  \\
\hline
$s_1$ (LR) &  0.2573  &  0.2621   & 0.0079 &\multirow{2}{*}{0.00091}\\
$s_2$ (LR) & 0.2724   & 0.2891    & 0.0374 &  \\
\hline
$s_1$ (RDBR) & 0.3321  &  0.3397  & 0.0218 &\multirow{2}{*}{16.5182} \\
$s_2$ (RDBR) & 0.2760  &  0.3012  & 0.0349 &  \\
\hline
$s_1$ (MMD*) & 0.2441 &  0.3450 & 0.2482 &\multirow{2}{*}{462.6}  \\
$s_2$ (MMD*) & 0.2454 &  0.3920 & 0.0490 &  \\
\end{tabular}
\end{center}
\end{table}

\section{The fourth simulated signal}

As a fourth and final example, we decomposed a three-component signal $f(t) = s_1(t) + s_2(t) + s_3(t)$, where
\begin{equation}\label{eq:s1_3}
s_1(t) = \cos(\phi_{11}(t)) + 0.5\cos(1.95 \phi_{11}(t) + 0.0001 \phi_{11}^2(t)),
\end{equation}
 with $\phi_{11}(t) = 2\pi 12 t + 2\pi 2 t^2$,
\begin{equation}\label{eq:s2_3}
\begin{aligned}
s_2(t) &= \cos(\phi_{21}(t)) + 0.75\cos(2.05 \phi_{21}(t)) \\
 & \quad + 0.25 \cos(2.95 \phi_{21}(t)),
\end{aligned}
\end{equation}
 with $\phi_{21}(t) = 2\pi 14 t + 2\pi t^2 + 2\pi t^3$, and
\begin{equation}\label{eq:s3_3}
s_3(t) = \sum_p 2000 (t-t_p) e^{-2\times10^5 (t-t_p)^2},
\end{equation}
with $t_p$ such that $\phi_{31}(t_p) = 2 \pi p, p\in \mathbb{Z}$ and $\phi_{31}(t) = 2 \pi 18 t + 2\pi 2t^2 + \cos(4 \pi t)$. As with the previous examples the signals are defined for $t\in[0,1]$, and sampled at 1000 Hz.

\begin{figure*}[t]
\begin{center}
\includegraphics[width=\textwidth]{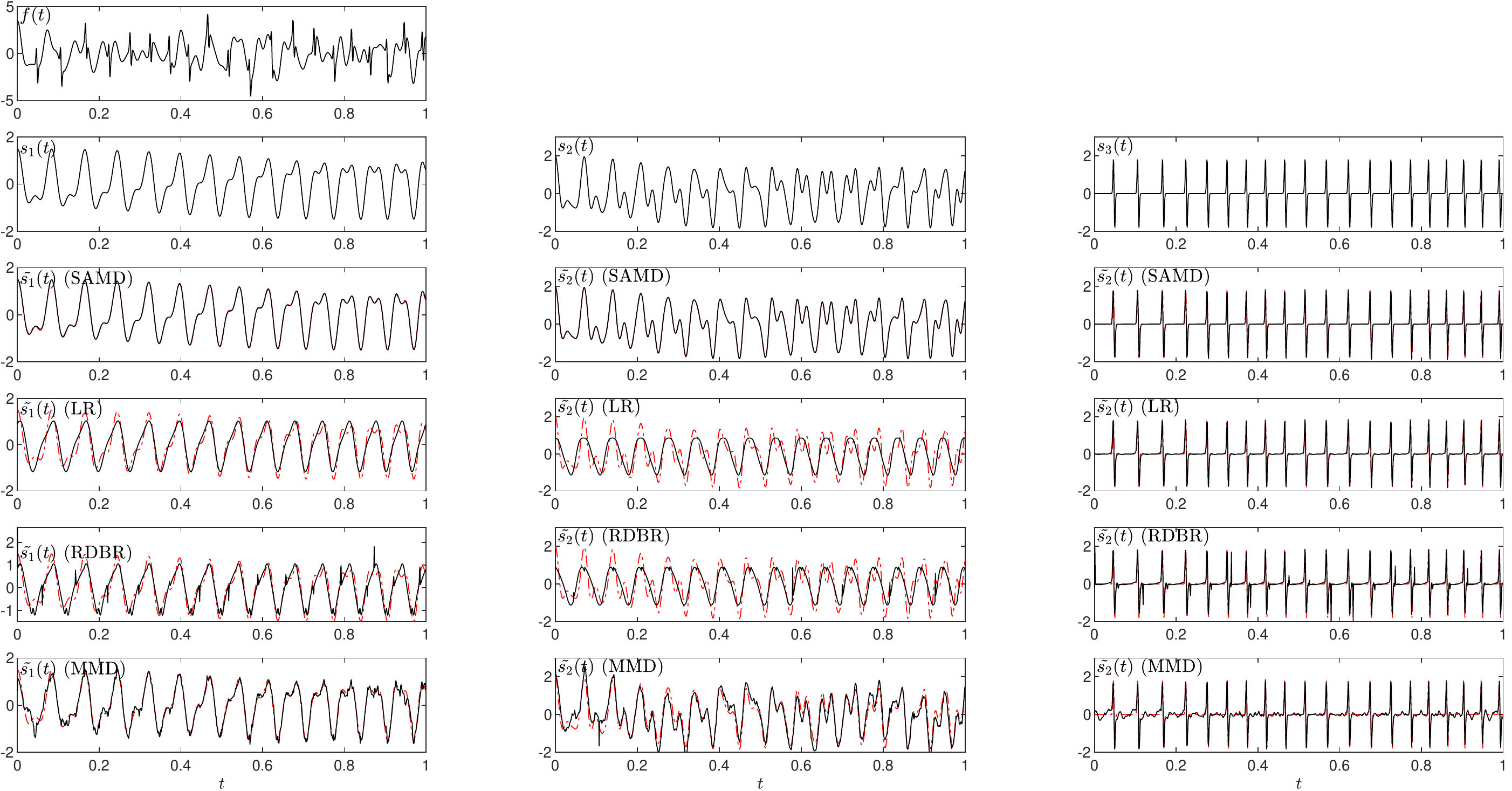}
\end{center}
\caption{\textbf{Fourth simulated signal (noiseless).} First row: analyzed signal $f(t)$ from Eqs. \eqref{eq:s1_3}, \eqref{eq:s2_3} and \eqref{eq:s3_3}. Second row: true components. Third row: extracted modes with our proposal SAMD. Fourth row: extracted components with LR. Fifth row: extracted components with RDBR. Sixth row: extracted components with MMD. (black solid line: estimated component; red dashed line: true components).}
\label{fig:03}
\end{figure*}

The obtained waveforms for the noiseless signal, via the four methods, can be found on Fig \ref{fig:03}. We used the parameters $D_1 = 2$, $D_2 = 3$, $D_3 = 20$ for both LR and SAMD methods, and $K = 3$ (this last one for the estimations of the phases for SAMD). Regarding $D_i$ parameters, these can be estimated by adapting trigonometric regression tools, where we minimize a criterion which is a function of the model order, looking for a trade-off between error and model order. Promising results of applications of these criteria on the wave-shape function model can be found in \cite{Ruiz2020}. In this case, as we said before, we assumed the knowledge of $\phi_{11}(t)$, $\phi_{21}(t)$ and $\phi_{31}(t)$. We measured the quality of modes recovery through RMSE, and the performance of the four methods can be found at Table I, with evident advantages for the SAMD method.

\begin{figure*}[t]
\begin{center}
\includegraphics[width=\textwidth]{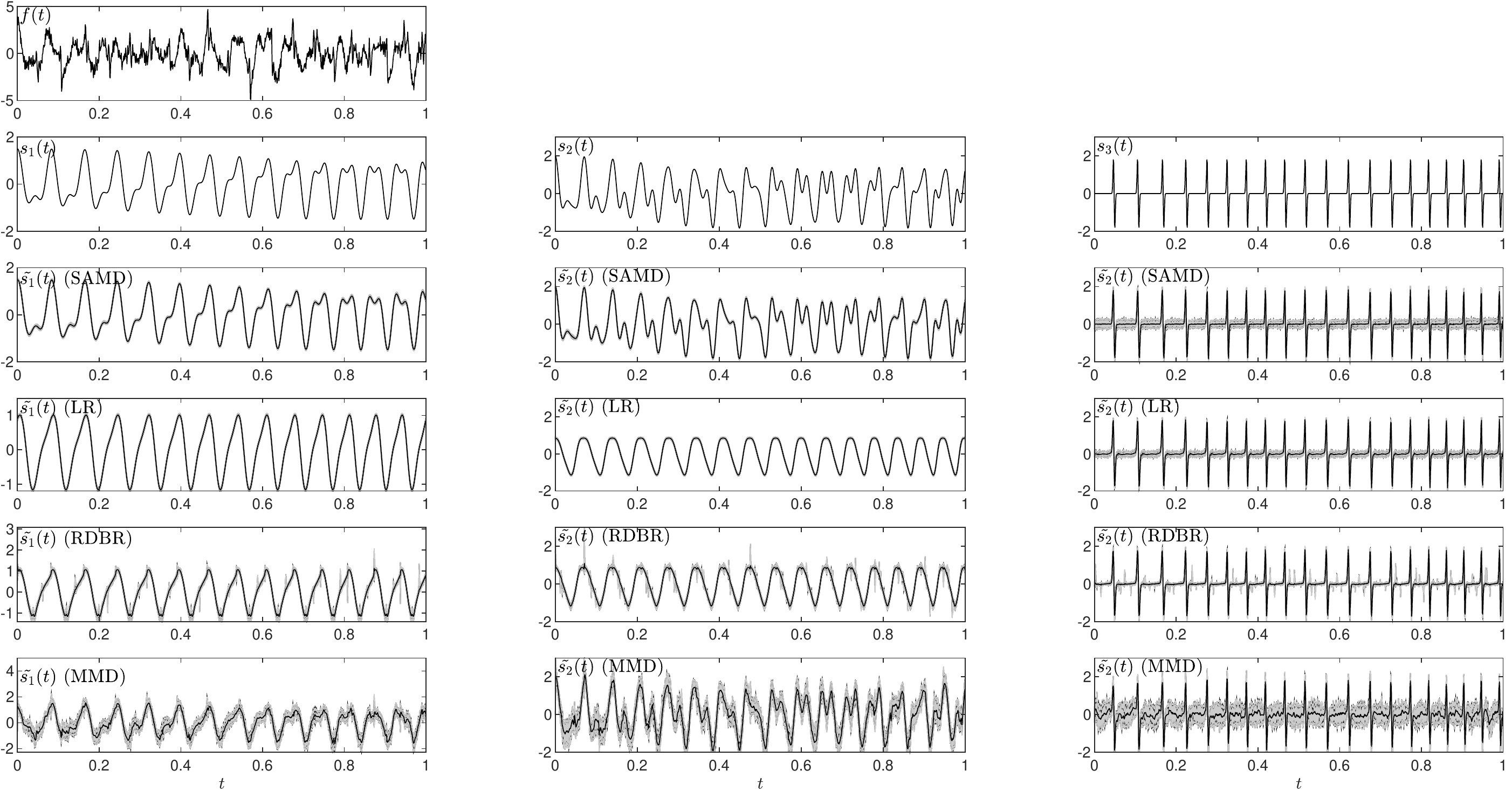}
\end{center}
\caption{\textbf{Fourth simulated signal (noisy signals at 10 dB; 100 realizations).} First row: a typical noisy example of signal $f(t)$ from Eqs. \eqref{eq:s1_3}, \eqref{eq:s2_3} and \eqref{eq:s3_3}. Second row: true components. Third row: extracted modes with our proposal SAMD. Fourth row: extracted components with LR. Fifth row: extracted components with RDBR. Sixth row: extracted components with MMD. (black solid line: mean estimated components; shaded gray area: 95\% confidence interval).}
\label{fig:03_ruido}
\end{figure*}

In order to test the robustness to noise of the different methods, we performed 100 decompositions of noisy versions of the signal at 10 dB. The noise is assumed to be Gaussian white. We present these results on Fig. \ref{fig:03_ruido}. We show, for each mode, the mean and the 95\% confidence interval. The means and standard deviations of the RMSE can be found at Table I, along with the mean computational times. These results evidence not only a better mode recovery performance for SAMD, but also a comparable computational load when compared to RDBR, and a significantly lower burden than MMD (at least one order of magnitude).

\begin{table}\label{table:03}
\caption{Errors and computation times for simulated signal from Eqs. \eqref{eq:s1_3}, \eqref{eq:s2_3}, and \eqref{eq:s3_3}. *Out of the 100 realizations, MMD converged on 96 occasions.}
\begin{center}
\begin{tabular}{lcccc}
\hline
   & Noiseless & \multicolumn{3}{c}{Noisy (10 dB; 100 realizations)} \\
             & RMSE   & mean(RMSE) & std(RMSE)& mean time (s) \\
\hline
$s_1$ (SAMD) & 0.0205 &  0.0372  & 0.0083 &\multirow{3}{*}{17.8319} \\
$s_2$ (SAMD) & 0.0036 & 0.0368 & 0.0102 &  \\
$s_3$ (SAMD) & 0.1416 & 0.1740 & 0.0056 &  \\
\hline
$s_1$ (LR) &  0.2989  &  0.3002   & 0.0012 &\multirow{3}{*}{0.0023}\\
$s_2$ (LR) & 0.5438   & 0.5446    & 0.0009 &  \\
$s_3$ (LR) & 0.1438   & 0.1650    & 0.0047 &  \\
\hline
$s_1$ (RDBR) & 0.3171  &  0.3119  & 0.0072 &\multirow{3}{*}{11.2058} \\
$s_2$ (RDBR) & 0.5470  &  0.5499  & 0.0181 &  \\
$s_3$ (RDBR) & 0.2121  &  0.1635  & 0.0165 &  \\
\hline
$s_1$ (MMD*) & 0.1620 &  0.2839 & 0.0138 &\multirow{3}{*}{522.5335}  \\
$s_2$ (MMD*) & 0.3302 &  0.2939 & 0.1435 &  \\
$s_3$ (MMD*) & 0.1480 &  0.3411 & 0.0137 &  \\
\end{tabular}
\end{center}
\end{table}

\begin{figure*}[t]
\begin{center}
\includegraphics[width=\textwidth]{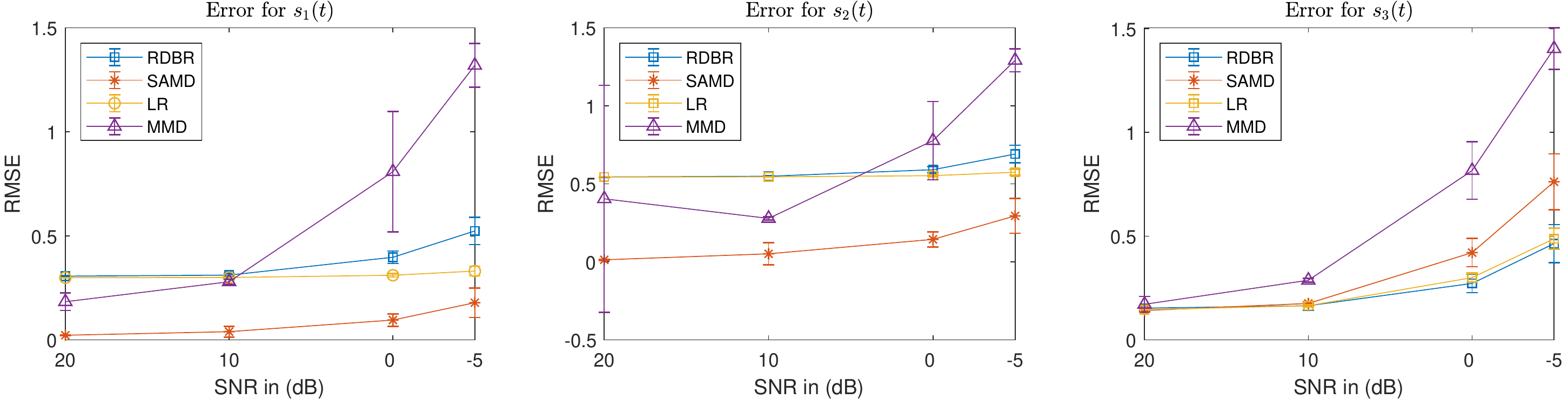}
\end{center}
\caption{\textbf{Errors for the fourth simulated signal (different SNRs; 50 realizations).} Left: mean errors and standard deviations for $s_1(t)$. Middle: mean errors and standard deviations for $s_2(t)$. Right: mean errors and standard deviations for $s_3(t)$. For MMD method, we considered only those decompositions that converged (between 44 and 49 times out of the 50 realizations, depending on the input SNR).}
\label{fig:03_varias}
\end{figure*}

We also tested noise robustness at different SNRs. We performed 50 decompositions of noisy versions of the signal at 20, 10, 0, and -5 dB, and computed the errors for $s_1(t)$, $s_2(t)$, and $s_3(t)$. Results can be appreciated on Fig. \ref{fig:03_varias}, where we present the mean of the RMSE and the standard deviations over the 50 decompositions. For $s_1(t)$, the results of SAMD are the best for all considered SNRs. The same is confirmed for $s_2(t)$, and as for $s_3(t)$ the performance results of SAMD, LR and RDBR are comparable when the SNR is equal to 20 and 10 dB. Out of the four analyzed methods, MMD seems to be the most sensitive to noise.

\subsection{Electrocardiogram signal from Fantasia Database}
We analyzed an ECG recording with noise and baseline wander as second real example. The analyzed recording belongs to the Fantasia Database \footnote{https://physionet.org/content/fantasia/1.0.0/} \cite{iyengar1996age,Physionet}.

\begin{figure*}[t]
\begin{center}
\includegraphics[width=\textwidth]{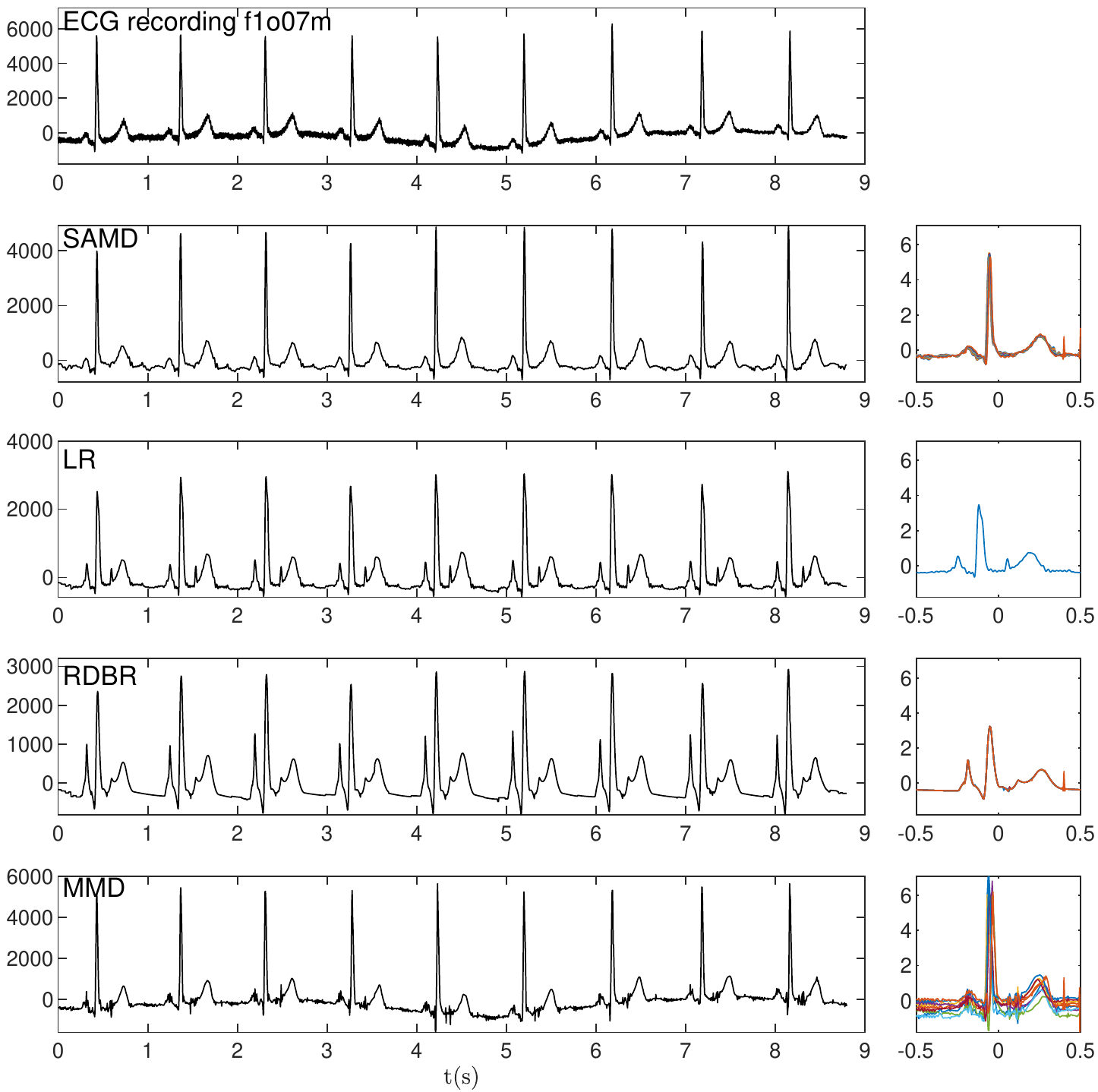}
\end{center}
\cprotect\caption{\textbf{ECG signal from Fantasia Database.} First row: 9 seconds of recording \verb"f1o07m". Second row: SAMD estimated component and waveforms. Third row: LR estimated component and waveform. Fourth row: RDBR estimated component and waveform. Fifth row: MMD estimated component and waveforms.}
\label{fig:ECG}
\end{figure*}

We used the four methods here for the missions of denoising and segmentation, and the results are presented in Fig. \ref{fig:ECG}. We assumed the signal has only one component. We used a Gaussian window with $\sigma = 0.15$, a maximum jump of 2 Hz and $\Delta = 2$Hz, for the estimation of amplitudes and phases. We used $D_1 = 40$ for LR and SAMD, and $K = 1$ for SAMD. The advantages of our SAMD method are evident here. Both LR and RDBR are not able to capture the fast QRS complexes, achieving a wider waveform. MMD was not able to remove the baseline wander, and the estimated waveforms contains an important amount of noise. SAMD, on the other hand, was able to remove the low frequency trend, with satisfactory waveforms that represents truly the different waves of the ECG cycle. The computational times were 36.9404s, 0.0263s, 4.5721s, and 421.9341s for SAMD, LR, RDBR and MMD respectively.

\end{document}